# PREDICTION MODELS FOR NETWORK-LINKED DATA

By Tianxi Li*, Elizaveta Levina† and Ji Zhu‡

*Department of Statistics, University of Michigan*

Prediction algorithms typically assume the training data are independent samples, but in many modern applications samples come from individuals connected by a network. For example, in adolescent health studies of risk-taking behaviors, information on the subjects' social network is often available and plays an important role through network cohesion, the empirically observed phenomenon of friends behaving similarly. Taking cohesion into account in prediction models should allow us to improve their performance. Here we propose a network-based penalty on individual node effects to encourage similarity between predictions for linked nodes, and show that incorporating it into prediction leads to improvement over traditional models both theoretically and empirically when network cohesion is present. The penalty can be used with many loss-based prediction methods, such as regression, generalized linear models, and Cox's proportional hazard model. Applications to predicting levels of recreational activity and marijuana usage among teenagers from the AddHealth study based on both demographic covariates and friendship networks are discussed in detail and show that our approach to taking friendships into account can significantly improve predictions of behavior while providing interpretable estimates of covariate effects.

**1. Introduction.** Advances in data collection and social media have resulted in network data being collected in many applications, recording relational information between units of analysis (Michell and West, 1996; Pearson and Michell, 2000; Pearson and West, 2003). This information is often collected along with more traditional covariates on each unit of analysis. One such case study we focus on in this paper is the survey data from the National Longitudinal Study of Adolescent Health (the AddHealth study) (Harris, 2009). AddHealth was a major national longitudinal study of students in grades 7-12 during the school year 1994-1995, after which three further follow-ups were conducted in 1996, 2001-2002, and 2007-2008. In the Wave I survey, all students in the sample completed in-school questionnaires, and a subsample completed a follow-up in-home interview with more detailed

*Partially supported by Rackham International Student Fellowship
†Supported by NSF grant DMS-1521551 and ONR grant N000141612910
‡Supported by NSF grant DMS-1407698 and NIH grant R01GM096194
*Keywords and phrases:* network cohesion, prediction, regression

questions. There are questions in both the in-school survey and the in-home interview asking students to name their friends (up to 10) so friendship networks connecting students can be constructed based on this information. In addition to the information about friends, the survey also asked hundreds of questions about various aspects of the students personal and school life, collecting information about age, gender, race, socio-economic status, health, academic achievement, etc.

There is a large body of work extending over decades on predicting a response variable of interest from such covariates, via linear or generalized linear models, survival analysis, classification methods, and the like, which typically assume the training samples are independent and do not extend to situations where the samples are connected by a network. There is also now a large body of work focusing on analyzing the network structure implied by the relational data alone, for example, detecting communities; see Goldenberg et al. (2010) and Abbe (2017) for reviews. The more traditional covariates, if used at all in such network analyses, are typically used to help analyze the network itself, e.g., find better communities (Binkiewicz, Vogelstein and Rohe, 2017; Newman and Clauset, 2016; Zhang et al., 2016). There has not been much focus on developing a general statistical framework for using network data in prediction, although there are methods available for specific applications (Wolf et al., 2009; Asur and Huberman, 2010; Vogelstein et al., 2013).

In the social sciences and especially in economics, on the other hand, there has been a lot of recent interest in causal inference on the relationship between a response variable and both covariates and network influences; see e.g., Shalizi and Thomas (2011) and references therein, and Manski (2013). While in certain experimental settings such inference is possible (Rand, Arbesman and Christakis, 2011; Choi, 2017; Phan and Airoldi, 2015), in most observational studies on networks establishing causality is substantially more difficult than in regular observational studies. While network cohesion (a generic term by which in this paper we mean linked nodes acting similarly) is a well known phenomenon observed in numerous social behavior studies (Fujimoto and Valente, 2012; Haynie, 2001; Christakis and Fowler, 2007), explaining it causally on the basis of observational data is very challenging. An excellent analysis of this problem can be found in Shalizi and Thomas (2011), showing that it is in general impossible to distinguish network cohesion resulting from homophily (nodes become connected because they act similarly) and cohesion resulting from contagion (behavior spreads from node to node through the links), and to separate that from the effect of node covariates themselves. However, making good predictions of node behavior is an easier task than causal inference, and is often all we need for practical purposes. Our goal in this paper is to take advantage of the network cohesion phenomenon in order to better predict a response variable associated with the network nodes, using both node covariates and network information. While we do not attempt to make causal inferences, we do focus on interpretable models where effects of individual variables can be explicitly estimated.

Using network information in predictive models has not yet been well studied. Most classical predictive models treat the training data as independently sampled from one common population, and, unless explicitly modeled, network cohesion violates virtually all assumptions that provide performance guarantees. More importantly, cohesion is potentially helpful in making predictions, since it suggests pooling information from neighboring nodes. In certain specific contexts, regression with dependent observations has been studied. For example, in econometrics, following the concepts initially discussed by Manski (1993), assuming some type of an auto-regressive model on the response variables is common, such as the basic autoregressive model in Bramoullé, Djebbari and Fortin (2009) and its variants including group interactions and group fixed effects (Lee, 2007). Such models assume specific forms of different types of network effects, namely, endogenous effects, exogenous effects and correlated effects, and most of this literature is focused on identifiability of such effects. In Bramoullé, Djebbari and Fortin (2009) and Lin (2010), these ideas were applied to the AddHealth data which we discuss in detail in Section 5. However, these methods have mainly been used to identify social effects defined within a very specific and difficult to verify model, without a focus on interpretability or good prediction performance. For instance, including neighbors' responses as covariates in linear regression makes interpretation of other covariate effects more difficult, and can make the distributional assumptions difficult to satisfy. This has been done carefully in spatial statistics literature, for example with the conditional autoregressive model (CAR) (Besag, 1974), but fitting these models typically requires MCMC and is very time-consuming. In addition, these methods do not extend easily beyond linear regression (for example, to generalized linear models and Cox's proportional hazard model).

Our approach is to introduce network cohesion using penalties built using the network information, and framing the problem as loss plus penalty; for simplicity, we will present the method for regression first, and then discuss extensions to general losses. At a high level, our network penalty parallels the ideas of fusion (Land and Friedman, 1997; Tibshirani et al., 2005). Fusion penalties generally shrink the difference between either coefficients or predictions that are expected to be similar. Fusion penalties based on a network of variables have been used in variable selection (Li and Li, 2008, 2010; Pan, Xie and Shen, 2010; Kim, Pan and Shen, 2013), but this line of work is not directly relevant here since we are interested in using the network of observations, not variables. However, our approach can be viewed as a regression version of the point estimation problem discussed in Sharpnack, Singh and Krishnamurthy (2013) and Wang et al. (2016). Alternatively, it can be viewed in a Bayesian framework, as regression with a Gaussian Markov random field prior over the network.

We show that our method gives consistent estimates of covariate effects and derive explicit conditions on when enforcing network cohesion in regression can be expected to perform better than ordinary least squares. In contrast to previous work, we assume no specific form for the cohesion effects and require no information about potential groups. We also derive a computationally efficient algorithm for implementing our approach, which is efficient for both sparse and dense networks, the latter with an extra sparsification step which we prove preserves the relevant network properties. To the best of our knowledge, this is the first proposal of a general prediction framework with network cohesion among the observations that is computationally feasible and can retain covariate interpretations as well as make out-of-sample predictions.

The rest of this paper is organized as follows. In Section 2, we introduce our approach in the setting of linear regression. We frame it as a penalized least squares problem which has a closed-form solution, and derive its Bayesian interpretation and connection to various other models. The idea is then extended to generalized linear models. Empirically, we show that our approach outperforms prediction without networks as well as an earlier modification intended to incorporate information from neighbors. Finite sample and asymptotic properties are discussed in Section 3. Brief simulation results demonstrating the theoretical bounds and comparisons to benchmarks are presented in Section 4. A detailed analysis and discussion of cohesion in the AddHealth data is presented in Section 5, where we apply our method to predict recreational activity and marijuana usage among teenagers. All algorithms in this paper are implemented in the R package **netcoh** (Li, Levina and Zhu, 2016), available on CRAN. Code for the examples in the paper can be found on the first author's webpage.

## 2. Regression with network cohesion.

2.1. *Set-up and notation.* We start from setting up notation. By default, all vectors are treated as column vectors. The data consist of $n$ ob-servations $(y_1, \boldsymbol{x}_1), (y_2, \boldsymbol{x}_2), \cdots, (y_n, \boldsymbol{x}_n)$, where $y_i \in \mathbb{R}$ is the response variable and $\boldsymbol{x}_i \in \mathbb{R}^p$ is the vector of covariates for observation $i$. We write $\boldsymbol{Y} = (y_1, y_2, \cdots, y_n)^T$ for the response vector, and $X = (\boldsymbol{x}_1, \boldsymbol{x}_2, \cdots, \boldsymbol{x}_n)^T$ for the $n \times p$ design matrix. We treat $X$ as fixed and assume its columns have been standardized to have mean 0 and variance 1. We also observe the network connecting the observations, $\mathcal{G} = (V, E)$, where $V = \{1, 2, \cdots, n\}$ is the node set of the graph, and $E \subset V \times V$ is the edge set. We represent the graph by its adjacency matrix $A \in \mathbb{R}^{n \times n}$, where $A_{uv} = 1$ if $(u, v) \in E$ and 0 otherwise. We assume there are no loops so $A_{vv} = 0$ for all $v \in V$, and the network is undirected, i.e., $A_{uv} = A_{vu}$. The (unnormalized) Laplacian of $\mathcal{G}$ is given by $L = D - A$, where $D = \mathrm{diag}(d_1, d_2, \cdots, d_n)$ is the degree matrix, with node degree $d_u$ defined by $d_u = \sum_{v \in V} A_{uv}$.

2.2. *Linear regression with network cohesion.* Cohesion is a vague term that can be interpreted in several ways depending on whether it refers to the network itself or both the network and additional covariates. Cohesion defined on the network alone can be reflected in various properties, such as local density, connectivity and community structure; we refer the readers to Chapter 4 of Kolaczyk (2009) for details. In the context of prediction on networks, which is our focus, two types of cohesion are commonly discussed: homophily (also known as assortative mixing) and contagion. Homophily means nodes similar in their characteristics tend to connect, with the implication of a causal direction from sharing individual characteristics to forming a connection. In contrast, contagion means that nodes tend to behave similarly to their neighbors, with a casual direction from having a connection to exhibiting similar characteristics. Distinguishing these two phenomena in an observational study without additional strong assumptions is not possible (Shalizi and Thomas, 2011). Nonetheless, both of these indicate a correlation between network connections and node similarities, observed empirically by many social behavior studies (Haynie, 2001; Pearson and West, 2003; Fujimoto and Valente, 2012), and that is all we need and assume in this paper. We use the generic term "cohesion" in order to cover both possibilities of homophily and contagion, which we do not need to distinguish.

The general cohesion penalty idea is simplest to present in the context of linear regression, so we start from this setting. Assume that

$$\boldsymbol{Y} = \boldsymbol{\alpha} + X\boldsymbol{\beta} + \boldsymbol{\epsilon} \tag{2.1}$$

where $\boldsymbol{\alpha} = (\alpha_1, \alpha_2, \cdots, \alpha_n)^T \in \mathbb{R}^n$ is the vector of individual node effects, and $\boldsymbol{\beta} = (\beta_1, \beta_2, \cdots, \beta_p)^T \in \mathbb{R}^p$ is the vector of regression coefficients. At this stage, no assumption on the distribution of the error $\boldsymbol{\varepsilon}$ is needed, but we assume $\mathbb{E}\boldsymbol{\epsilon} = \mathbf{0}$ and $\text{Var}(\boldsymbol{\epsilon}) = \sigma^2 I_n$, where $I_n$ is the $n \times n$ identity matrix. For simplicity, we further assume that $n > p$ and $X^T X$ is invertible. If $p > n$ and this is not the case, the usual penalties on $\boldsymbol{\beta}$, such as a lasso and ridge, can be applied; our focus here, however, is on regularizing the individual effects, and so we will not focus on additional regularization on $\boldsymbol{\beta}$ that may be necessary.

Including the individual node effects $\boldsymbol{\alpha}$ instead of a common shared intercept turns out to be key to incorporating network cohesion. In general $\boldsymbol{\alpha}$ and $\boldsymbol{\beta}$, which add up to $n + p$ unknown parameters, cannot be estimated from $n$ observations without additional assumptions. One well-known example of such assumptions is the simple fixed effects model (see e.g. Searle, Casella and McCulloch (2009)), when $n$ samples come from $K$ known groups (typically $K \ll n$), and within each group individuals share a common intercept. Here, we regularize the problem through a network cohesion penalty on $\boldsymbol{\alpha}$ instead of making explicit assumptions about any structure in $\boldsymbol{\alpha}$.

The regression with network cohesion (RNC) estimator we propose is defined as the minimizer of the objective function

$$L(\boldsymbol{\alpha}, \boldsymbol{\beta}) = \|\boldsymbol{Y} - X\boldsymbol{\beta} - \boldsymbol{\alpha}\|^2 + \lambda \boldsymbol{\alpha}^T L \boldsymbol{\alpha}, \tag{2.2}$$

where $\|\cdot\|$ is the $L_2$ vector norm and $\lambda > 0$ is a tuning parameter. An equivalent and more intuitive form of the penalty, which follows from a simple property of the graph Laplacian, is

$$\boldsymbol{\alpha}^T L \boldsymbol{\alpha} = \sum_{(u,v) \in E} (\alpha_u - \alpha_v)^2. \tag{2.3}$$

Thus, we penalize differences between individual effects of nodes connected by an edge in the network. We call this term the *cohesion penalty* on $\boldsymbol{\alpha}$. We assume that the effect of covariates $X$ is the same across the network; as with any linear regression, two nodes with similar covariates will have similar values of $\boldsymbol{x}^T\boldsymbol{\beta}$, and the cohesion penalty ensures the neighboring nodes have similar individual effects $\alpha$. Note that this is different from imposing network homophily (which would require nodes with similar covariates to be more likely to be connected).

The minimizer of (2.2) can be computed explicitly (if it exists) as

$$\hat{\boldsymbol{\theta}} = (\hat{\boldsymbol{\alpha}}, \hat{\boldsymbol{\beta}}) = (\tilde{X}^T \tilde{X} + \lambda M)^{-1} \tilde{X}^T \boldsymbol{Y}. \tag{2.4}$$

Here, $\tilde{X} = (I_n, X)$ and

$$M = \begin{bmatrix} L & 0_{n \times p} \\ 0_{p \times n} & 0_{p \times p} \end{bmatrix}$$

where $0_{a\times b}$ is an $a \times b$ matrix of all zeros. The estimator exists if $\tilde{X}^T\tilde{X} + \lambda M$ is invertible. Note that

$$\tilde{X}^T\tilde{X} + \lambda M = \begin{bmatrix} I_n + \lambda L & X \\ X^T & X^T X \end{bmatrix}, \tag{2.5}$$

so it is positive definite if and only if the Schur complement $I_n + \lambda L - X(X^TX)^{-1}X^T = P_{X^\perp} + \lambda L$ is positive definite. From (2.3), we can see that $L$ is positive semi-definite but singular since $L\mathbf{1}_n = 0$ where $\mathbf{1}$ is the vector of all ones, and thus in principle the estimator may not be computable. In Section 3, we will give an interpretable theoretical condition for the estimator to exist. In practice, a natural solution is to ensure numerical stability by replacing $L$ with the regularized Laplacian $L + \gamma I$, where $\gamma$ is a small positive constant. Then the estimator always exists, and in fact the regularized Laplacian may better represent certain network properties, as discussed by Chaudhuri, Graham and Tsiatas (2012); Amini et al. (2013); Le, Levina and Vershynin (2017) and others. The resulting penalty is

$$\sum_{(u,v)\in E} (\alpha_u - \alpha_v)^2 + \gamma \sum_v \alpha_v^2, \tag{2.6}$$

which one can also interpret as adding a small ridge penalty on $\alpha$ for numerical stability.

Remark 1. The penalty (2.6) suggests a natural baseline comparison for our model which can be used to assess whether cohesion is in fact present in the data. If the graph has no edges. i.e., no information about network connections is available, the penalty (with $\gamma = 1$) reduces to a ridge penalty on the individual effects $\alpha$. The parameter estimates are then obtained by minimizing

$$L_n(\boldsymbol{\alpha}, \boldsymbol{\beta}) = \|\boldsymbol{Y} - X\boldsymbol{\beta} - \boldsymbol{\alpha}\|^2 + \lambda\|\boldsymbol{\alpha}\|^2 \ . \tag{2.7}$$

We call this the *null model* for RNC, as it still incorporates individual node effects which in themselves can improve performance compared to OLS with a common intercept. As discussed later in Section 2.4 and 2.7, this null model can also be viewed as a random effects model with i.i.d Gaussian intercepts. Comparing the fit of the null model to that of RNC can in fact provide qualitative evidence of cohesion. For linear regression, the null model can improve the fit to training data, but it gives exactly the same estimate of $\boldsymbol{\beta}$ as the OLS (Lemma 3 in Appendix A), and thus cannot improve predictions on test data, since without network information individual effects on test data cannot be estimated; see more on this in Section 4.

Remark 2. A possible alternative to our cohesion penalty is the network lasso penalty, $\sum_{(u,v)\in E} |\alpha_u - \alpha_v|$ (Hallac, Leskovec and Boyd, 2015). However, this penalty introduces piecewise constants on the network, a rather stronger assumption than we make about cohesion which may not be always realistic. It is also much more computationally demanding, requiring a sophisticated algorithm and implementation even for moderate size networks.

Remark 3. It is also possible to assume different but cohesive covariate effects $\boldsymbol{\beta}$ for each individual, which can be implemented in exactly the same way as our idea of the individual intercepts $\alpha$. As usual, there is a tradeoff between including more parameters for better fit and parsimony of the model. We set $\boldsymbol{\beta}$ to be shared among all individual to represent the universal treatment effect, which seems to be reasonable and easy to interpret in many situations.

2.3. *Network cohesion for general loss functions.* The RNC methodology extends naturally to generalized linear models and many other regression or classification models, such as Cox's proportional hazard model (Cox, 1972) for survival analysis, and support vector machines (Vapnik, 2013) for classification using the formulation of Wahba et al. (1999). Here we will explicitly write out two extensions, to generalized linear models (GLMs) and Cox's model. For any GLM with a link function $\phi(\mathbb{E}\boldsymbol{Y}) = X\boldsymbol{\beta} + \boldsymbol{\alpha}$, where $\boldsymbol{\alpha} \in \mathbb{R}^n$ are the individual effects, suppose the log-likelihood (or partial log-likelihood) function is $\ell(\boldsymbol{\alpha}, \boldsymbol{\beta}; X, \boldsymbol{Y})$. Then if the observations are linked by a network, to induce network cohesion one can fit the model by maximizing the penalized likelihood

$$\ell(\boldsymbol{\alpha} + X\boldsymbol{\beta}; \boldsymbol{Y}) - \lambda \boldsymbol{\alpha}^T (L + \gamma I)\boldsymbol{\alpha}. \tag{2.8}$$

When $\ell$ is concave in $\boldsymbol{\alpha}$ and $\boldsymbol{\beta}$, which is the case for exponential families, the optimization problem can be solved via Newton-Raphson or another appropriate convex optimization algorithm. Note that the quadratic approximation to (2.8) is the quadratic approximation to the log-likelihood plus the penalty, and thus the problem can be efficiently solved by iteratively reweighed linear regression with network cohesion, just like the GLM is fitted by iteratively reweighed least squares. The ridge penalty term $\gamma I$ helps with numerical stability and for logistic regression avoids fitted probabilities of 0 and 1 for isolated nodes, which may cause the iterative algorithm to diverge; as discussed in the previous section, adding this term to the Laplacian also improves its representation of the underlying network structure.

RNC can be similarly generalized to Cox's proportional hazard model (Cox, 1972). In this setting, we observe times until some event occurs, called survival times, which may be censored (unobserved) if the event has not occurred for a particular node. Cox's model assumes the hazard function $h_v(y)$ for each individual $v$ is

$$h_v(y) = h_0(y) \exp(\boldsymbol{x}_v^T \boldsymbol{\beta}), v \in V,$$

where $y$ is the survival time, $\boldsymbol{x}_v$ is the vector of $p$ observed covariates for individual $v$, $\boldsymbol{\beta} \in R^p$ is the coefficient vector and $h_0$ is an unspecified baseline hazard function. When we have observations connected by a network, we can model the individual effects and then encourage network cohesion. Thus we will assume the hazard for each node $v$ is given by

$$h_v(y) = h_0(y) \exp(\boldsymbol{x}_v^T \boldsymbol{\beta} + \alpha_v), v \in V, \tag{2.9}$$

where $\alpha_v$ is the individual effect of node $v$. The appropriate loss function in terms of the parameters $\boldsymbol{\theta} = (\boldsymbol{\alpha}, \boldsymbol{\beta})$ is the partial log-likelihood

$$\ell(\boldsymbol{\theta}; \boldsymbol{y}) = \sum_v \delta_v \left[ \boldsymbol{x}_v^T \boldsymbol{\beta} + \alpha_v - \log \big( \sum_{u: y_u \geq y_v} \exp(\boldsymbol{x}_u^T \boldsymbol{\beta} + \alpha_u) \big) \right] \tag{2.10}$$

where $y_v$ is the observed survival time for node $v$, and $\delta_v$ is the censoring indicator, which is 0 if the observation is right-censored and 1 otherwise. Note that the partial log-likelihood is invariant under a shift in $\boldsymbol{\alpha}$ since such a shift can always be absorbed into $h_0$. Thus for identifiability, we require $\sum \alpha_v = 0$. For fixed covariates $\boldsymbol{x}_v$, $\alpha_v$ is the individual deviation from the population average log-hazard. The sum-to-zero constraint can be automatically enforced by replacing the network Laplacian $L$ in the network cohesion penalty with its regularized version $L + \gamma I$, or equivalently adding a ridge penalty on $\alpha$'s. Thus we maximize the following objective function, adding a regularized cohesion penalty to the partial log-likelihood:

$$\ell(\boldsymbol{\theta}) - \lambda \boldsymbol{\alpha}^T (L + \gamma I) \boldsymbol{\alpha}.$$

2.4. *A Bayesian interpretation.* The RNC estimator can also be framed as a Bayesian regression model. Consider the model

$$\boldsymbol{Y} | \boldsymbol{\alpha}, \boldsymbol{\beta} \sim \mathcal{N}(\boldsymbol{\alpha} + X\boldsymbol{\beta}, \sigma^2 I), \quad \boldsymbol{\beta} \sim \pi_{\boldsymbol{\beta}}(\phi), \quad \boldsymbol{\alpha} \sim \pi_{\boldsymbol{\alpha}}(\Phi),$$

where $\pi_{\boldsymbol{\beta}}(\phi)$ is the prior for $\boldsymbol{\beta}$ with hyperparameter $\phi$, $\pi_{\boldsymbol{\alpha}}(\Phi)$ is the prior for $\boldsymbol{\alpha}$ with hyperparameter $\Phi$, and $\sigma^2$ is assumed to be known. Suppose we take $\pi_{\boldsymbol{\beta}}(\phi)$ to be the non-informative Jeffrey's prior, reflecting lack of prior knowledge about the coefficients, and set $\pi_{\boldsymbol{\beta}}(\phi) \propto 1$. For $\boldsymbol{\alpha}$, assume a Gaussian Markov random field (GMRF) prior $\pi_{\boldsymbol{\alpha}} = \mathcal{N}_{\mathcal{G}}(\mathbf{0}, \Phi)$, where $\Phi = \Omega^{-1} = \zeta^2(L + \gamma I)^{-1}$. Note that when $\gamma = 0$, $\Omega$ is not invertible, and $\pi_{\boldsymbol{\alpha}}$ is an improper prior called intrinsic GMRF (Rue and Held, 2005).

If the posterior modes are used as the estimators for $\boldsymbol{\alpha}$ and $\boldsymbol{\beta}$, then this is equivalent to (2.2) with $\lambda = \sigma^2/\zeta^2$ and the Laplacian replaced by the regularized Laplacian $L + \gamma I$. Thus the estimator of (2.2) is the Bayes estimator with the improper intrinsic GMRF prior over the network on $\boldsymbol{\alpha}$. Note that this Bayesian interpretation is also valid for the generalized linear models.

2.5. *Prediction and choosing the tuning parameter.* To compute fitted values on the training data (in-sample prediction), we simply use $\hat{\boldsymbol{\alpha}} + X\hat{\boldsymbol{\beta}}$. The out-of-sample prediction task in this setting is to make predictions on a group of new subjects whose covariates as well as network connections (but not responses) become available after the model is fitted on training data. Since we have a different $\alpha_v$ for each node $v$, predicted individual effects are needed for new samples. Suppose we have a total of $n$ training samples and $n'$ test samples, resulting in a new network with $n + n'$ nodes where the first $n$ nodes are from training and the last $n'$ are the test nodes. Write the associated Laplacian as

$$L' = \begin{bmatrix} L_{11} & L_{12} \\ L_{21} & L_{22} \end{bmatrix},$$

where $L_{11}$ corresponds to the original $n$ training samples and $L_{22}$ corresponds to the $n'$ test samples. Similarly write the individual effect vector as $(\boldsymbol{\alpha}_1, \boldsymbol{\alpha}_2)$, where $\boldsymbol{\alpha}_1 = \hat{\boldsymbol{\alpha}}$ is estimated from training data, and $\boldsymbol{\alpha}_2$ needs to be predicted.

To take advantage of cohesion, we predict $\boldsymbol{\alpha}_2$ by minimizing the overall cohesion penalty, letting

$$\hat{\boldsymbol{\alpha}}_2 = \arg\min_{\boldsymbol{\alpha}_2} (\hat{\boldsymbol{\alpha}}, \boldsymbol{\alpha}_2)^T L' (\hat{\boldsymbol{\alpha}}, \boldsymbol{\alpha}_2) \ .$$

This gives

$$\hat{\boldsymbol{\alpha}}_2 = -L_{22}^{-1} L_{21} \hat{\boldsymbol{\alpha}}.$$

This corresponds to a supervised prediction setting, our focus in this paper, which assumes only the training data are available at the time of fitting. Our method can also be used in a semi-supervised setting, where the entire network is available at the time of training. In this case, the cohesion penalty at the fitting stage can include all the individual effects for all data points and the entire network so $\boldsymbol{\alpha}_1$ and $\boldsymbol{\alpha}_2$ are jointly optimized simultaneously.

The tuning parameter $\lambda$ can be selected by cross-validation. Randomly splitting or sampling from a network is not straightforward; however, we found that the usual "naive" cross-validation finds very good tuning parameters for our method, perhaps because it is fundamentally a regression problem and we are not attempting to make any inferences about the structure of the network. We tune using regular 10-fold cross-validation, randomly splitting the samples into 10 folds, leaving each fold out in turn, and training the model using the remaining nine folds and the corresponding induced subnetwork. The cross-validation error is computed as the average of the prediction errors on the fold that was left out, and the tuning parameter is picked to minimize the cross-validation error.

2.6. *An efficient computation strategy.* Computing the estimator (2.4) involves solving a $(n+p)\times(n+p)$ linear system so a naive implementation would require $O((n+p)^3)$ operations. For GLMs, such a system has to be solved in each Newton step. This computational burden can be reduced significantly by taking advantage of the fact that most networks in practice have sparse adjacency matrices as well as sparse Laplacians, which allows for using block elimination. A general description of this strategy can be found in many standard texts (see e.g. Boyd and Vandenberghe (2004), Ch. 4). Here we give the details in our setting.

The linear system we need to solve is

$$(\tilde{X}^T\tilde{X}+\lambda M)\boldsymbol{a}=\boldsymbol{b}.$$

From (2.5), we can rewrite this system with the following block structure:

$$\begin{bmatrix} I+\lambda L & X \\ X^T & X^TX \end{bmatrix}\begin{bmatrix}\boldsymbol{a}_1\\ \boldsymbol{a}_2\end{bmatrix}=\begin{bmatrix}\boldsymbol{b}_1\\ \boldsymbol{b}_2\end{bmatrix}.$$

The top row gives

$$(I+\lambda L)\boldsymbol{a}_1=(\boldsymbol{b}_1-X\boldsymbol{a}_2)$$

and substituting this into the bottom row, we have

$$(X^TX-X^T(I+\lambda L)^{-1}X)\boldsymbol{a}_2=\boldsymbol{b}_2-X^T(I+\lambda L)^{-1}\boldsymbol{b}_1.$$

Note that $I+\lambda L$ is a symmetric diagonal dominant (SDD) matrix, and is sparse most of the time in practice, so $(I+\lambda L)^{-1}\boldsymbol{b}_1$ and $(I+\lambda L)^{-1}X$ can be efficiently computed (Koutis, Miller and Peng, 2010; Cohen et al., 2014). The cost of this step is roughly $O(p(n+2|E|)(\log n)^{1/2})$, where $|E|$ is the number of edges in the network and $c$ is some absolute constant. The cost of the remaining computations is dominated by the cost of inverting the $p\times p$ matrix $X^T X - X^T(I + \lambda L)^{-1}X$, which is of the same order as the cost of solving a standard least squares problem.

When $A$ and $L$ are dense matrices, with $|E| = O(n^2)$, the strategy above has the cost of $O(pn^2((\log n)^{1/2})$, which is still better than naively solving the system, but we do not gain anything from block elimination unless $L$ is sparse. However, we can first apply a graph sparsification algorithm to $A$ and use the sparsified $A^*$ as input for RNC. For instance, the algorithm of Spielman and Teng (2011) can find $A^*$ with $O(\epsilon^{-2} n \log n)$ edges at the cost of $O(|E| \log^2 n)$ operations such that its sparsified Laplacian $L^*$ satisfies

$$(1 - \epsilon)L \preceq L^* \preceq (1 + \epsilon)L,$$

for a given constant $\epsilon > 0$. After this sparsification step, the complexity of solving the linear system reduces to to $O(pn \log^c n)$ for $c \leq 3$. In Section 3, we will provide theoretical guarantees for the accuracy of the RNC estimator based on $L^*$ compared to that based on $L$.

When the number of edges is on the order of $O(n^2)$, the sparsification step itself has complexity of $O(n^2 \log^c n)$, which is not necessarily cheaper than directly solving the original dense linear system using the SDD property. However, the advantage of sparsification becomes obvious when one has to iteratively solve the linear system for the GLM or Cox's model, and/or compute a solution path for a sequence of $\lambda$ values. In such situations, sparsificaiton only has to be done once and the average complexity of solving the linear system can be close to $O(n \log^c n)$ for the whole estimation procedure. Details of complexity calculations for the RNC are given in Appendix B; a more comprehensive discussion of the computational trade-off of sparsification can be found in Sadhanala, Wang and Tibshirani (2016).

### 2.7. *Connection to other models.*

*Fixed group effects models.* The fixed effects regression model with subjects divided into groups is a special case of RNC. If the graph $\mathcal{G}$ represents the groups as cliques (everyone within the same group is connected), there are no connections between groups, and we let $\lambda \to \infty$, then all nodes in one group will share a common intercept.

*Mixed effects models..* A mixed model, like ours, has individual effects viewed as random ($\boldsymbol{\alpha}$) and fixed covariate effects ($\boldsymbol{\beta}$), but no network effects. Our null model is a standard mixed model. The Bayesian interpretation of our method suggests we are inducing correlations between the random effects, $\boldsymbol{\alpha} \sim \mathcal{N}_{\mathcal{G}}(0, \Phi)$. The estimator (2.4) is then the mixed model equation in Henderson (1953) for estimating fixed effects and predicting random effects simultaneously (see Searle, Casella and McCulloch (2009)). However, the framework of mixed models requires stronger assumptions on the form of variance components. Moreover, (generalized) mixed models are not designed for predictions conceptually, and we will show in the simulation study as well as theoretically in Lemma 3 in Appendix A that the null model is not able to improve on out-of-sample predictions.

*Spatial models.* In spatial statistics, data points are typically indexed by their locations. A weight matrix $A$ can be computed as a function of distance between locations and can be used as a weighted analogue of our network adjacency matrix. This leads to natural connections between RNC and methods used in spatial statistics. In particular, ignoring the covariates $X$, RNC reduces to the Laplacian smoothing point estimation procedure in Sharpnack, Singh and Krishnamurthy (2013) and Wang et al. (2016), which is equivalent to krigging in spatial statistics (Cressie, 1990). It has been shown that a class of semi-supervised learning methods based on Laplacian smoothing can be viewed as "graph krigging" (Xu, Dyer and Owen, 2010) . From this perspective, RNC can be viewed as a generalization of graph krigging of Xu, Dyer and Owen (2010) to incorporate covariates and general loss functions. With covariates $X$ included, the Bayesian interpretation of RNC assumes the same Gaussian Markov random field distribution for $\alpha$ as the conditional autoregressive model (CAR) (Besag, 1974) and its GLM generalization (Chapter 9 of Waller and Gotway (2004)) assume for errors in spatial regression. However, $\zeta^2$ and $\sigma^2$ in our Bayesian interpretation are treated as parameters in the CAR, while $\lambda = \sigma^2/\zeta^2$ is treated as a tuning parameter in RNC. Further, the CAR model is fitted either by maximum likelihood involving computationally expensive integration steps, or by posterior inference via Markov chain Monte Carlo after assuming a full Bayesian model with additional priors on $\boldsymbol{\beta}$ and $\zeta^2$, etc. Both ways require much heavier computations than RNC, especially for GLM where the Gaussian Markov random field is no longer the conjugate prior. More importantly, CAR models cannot be applied to general loss functions that are not a well-defined likelihood, for example, for Cox's model and SVM. Also, CAR models suffer from conceptual difficulties in making out-of-sample predictions (Waller and Gotway, 2004). In contrast, RNC provides a universal strategy under general loss functions and comes with a natural out-of-sample predictor, discussed in Section 2.5.

*Manifold embeddings.* Our Laplacian-based penalty has connections to the large literature on manifold embeddings and semi-supervised learning. The general task of manifold embeddings is to embed data points, typically observed in some high-dimensional space equipped with a potentially non-Euclidean similarity measure, into a low-dimensional Euclidean space, while preserving dissimilarity between the points as much as possible. Finding the "right" embedding space is expected to help with downstream analysis tasks, such as visualization (Tenenbaum, De Silva and Langford, 2000) or clustering (Shi and Malik, 2000). Perhaps the algorithm most closely related to ours is Laplacian Eigenmaps (Belkin and Niyogi, 2003), which proposed using $k$ eigenvectors of the constructed graph Laplacian $L$ corresponding to the smallest eigenvalues as the Euclidean embedding of the graph in order to obtain a low-dimensional representation of the data, and its kernel version with a regularization penalty (Belkin, Niyogi and Sindhwani, 2006). There are multiple semi-supervised learning approaches to prediction on manifolds, where it is assumed that all the similarities (corresponding to the network in our case) are observed but only some of the data points are labelled (Zhou et al., 2004; Zhou, Huang and Schölkopf, 2005). Later out-of-sample extensions (Bengio et al., 2004; Cai, He and Han, 2007; Vural and Guillemot, 2016) were developed by assuming the embedding coordinates take certain specific forms as functions of the original data points, and in general the manifold literature relies on an underlying Euclidean space where distance and smoothness are well defined, an assumption we do not make.

Supervised manifold embeddings have also been proposed when class labels are available in training data, including for the Laplacian Eigenmaps (Yang, Sun and Zhang, 2011; Raducanu and Dornaika, 2012; Vural and Guillemot, 2016). The basic idea is to learn a low-dimensional embedding of the data that also corresponds to a good separation of classes, and then use the coordinates in this embedding as predictors instead of the original variables. For general response variables instead of class labels, there is no supervised variant of Laplacian Eigenmaps as far as we are aware. More importantly, the embedding coordinates are typically complicated implicit functions of all the variables, and their coefficients cannot be interpreted in any meaningful way. Our method, on the other hand, has the original variables as predictors in the model (and nothing else), and thus their regression coefficients are readily interpretable.

**3. Theoretical properties of the RNC estimator.** Recall the RNC estimator is given by

$$\hat{\boldsymbol{\theta}} = (\tilde{X}^T\tilde{X} + \lambda M)^{-1}\tilde{X}^T\mathbf{Y}, \tag{3.1}$$

where

$$M = \begin{bmatrix} L & 0 \\ 0 & 0 \end{bmatrix}.$$

We continue to assume that $X$ has centered columns and full column rank. Intuitively, we expect the network cohesion effect to improve prediction only when the network provides "new" information that is not already contained in the predictors $X$. We formalize this intuition in the following assumption:

ASSUMPTION 1. For any $\boldsymbol{u} \neq 0$ in the column space of $X$, $\boldsymbol{u}^T L \boldsymbol{u} > 0$.

This natural and fairly mild assumption is enough to ensure the existence of the RNC estimator. Write $\text{col}(X)$ for the linear space spanned by columns of $X$ and $\text{col}(X)^\perp$ for its orthogonal complement. Then the projection matrix onto $\text{col}(X)^\perp$ is $P_{X^\perp} = I_n - P_X$, where $P_X = X(X^T X)^{-1} X^T$. Write $\lambda_{\min}(M)$ for the minimum eigenvalue of any matrix $M$. Then we have the following lemma:

PROPOSITION 1. Whenever $\lambda > 0$, we have $0 \leq \nu = \lambda_{\min}(P_{X^\perp} + \lambda L) \leq 1$. Under Assumption 1 the RNC estimator (3.1) exists.

Lemma 1 in the Appendix shows that when the network is connected and $X$ is centered, the RNC estimator always exists since in a connected graph, $L$ has rank $n-1$, and an eigenvector $\mathbf{1}$.

THEOREM 3.1. *Under Assumption 1, the RNC estimator* $\hat{\boldsymbol{\theta}} = (\hat{\boldsymbol{\alpha}}, \hat{\boldsymbol{\beta}})$ *defined by* (3.1) *satisfies*

$$\text{MSE}(\hat{\boldsymbol{\alpha}}) \leq \frac{\lambda^2}{\nu^2}\|L\boldsymbol{\alpha}\|^2 + \frac{n}{\nu}\sigma^2, \tag{3.2}$$

$$\text{MSE}(\hat{\boldsymbol{\beta}}) \leq \frac{\lambda^2}{\nu^2 \mu}\|L\boldsymbol{\alpha}\|^2 + \sigma^2(\frac{1}{\nu} + 1)\text{tr}((X^T X)^{-1}), \tag{3.3}$$

$$\mathbb{E}\|\hat{\boldsymbol{Y}} - \mathbb{E}\boldsymbol{Y}\|^2 \leq \frac{\lambda^2}{\nu}\|L\boldsymbol{\alpha}\|^2 + \sigma^2\|S_\lambda\|_F^2, \tag{3.4}$$

*where the minimum eigenvalue of* $X^T X$ *is denoted by* $\mu$ *and* $\|S_\lambda\|_F$ *is the Frobenius norm of the shrinkage matrix* $S_\lambda = \tilde{X}(\tilde{X}^T \tilde{X} + \lambda L)^{-1} \tilde{X}^T$. *In particular, when* $\|L\boldsymbol{\alpha}\| = 0$, *and therefore* $\boldsymbol{\alpha}$ *is constant over each connected component of the network, RNC is unbiased.*

The proof is given in the Appendix where the expressions for exact errors are also available. Theorem 3.1 applies to any fixed $n$. The asymptotic results as the size of the network $n$ grows are presented next in Theorem 3.2. We add the subscript $n$ to previously defined quantities to emphasize the asymptotic nature of this result.

THEOREM 3.2. *If Assumption 1 holds,* $\mu_n = O(n)$, $\|L_n\boldsymbol{\alpha}_n\|^2 = o(n^c)$ *for some constant* $c < 1$, *and there exists a sequence of* $\lambda_n$ *and a constant* $\rho > 0$ *such that* $\liminf_n \nu_n > \rho$, *then*

$$\mathrm{MSE}(\hat{\boldsymbol{\beta}}) \leq O(\lambda_n^2 n^{-(1-c)}) + O(n^{-1}).$$

*Therefore if* $\lambda_n^2 = o(n^{1-c})$, $\hat{\boldsymbol{\beta}}$ *is an* $L_2$*-consistent estimator of* $\boldsymbol{\beta}$.

REMARK 4. Note that the quantity $L\boldsymbol{\alpha}$ appearing in the assumptions is the gradient of the cohesion penalty with respect to $\boldsymbol{\alpha}$, $\nabla_{\boldsymbol{\alpha}}\boldsymbol{\alpha}^T L\boldsymbol{\alpha} = 2L\boldsymbol{\alpha}$. We call $L\boldsymbol{\alpha}$ the cohesion gradient. In physics, cohesion gradient is used to measure heat diffusion on graphs when $\boldsymbol{\alpha}$ is a heat function:

$$(L\boldsymbol{\alpha})_v = |N(v)|\left(\alpha_v - \frac{\sum_{u\in N(v)}\alpha_u}{|N(v)|}\right).$$

where $N(v)$ is the set of neighbors of $v$ defined by the graph. Thus $\|L\boldsymbol{\alpha}\|$ represents the difference between nodes' individual effects and the average of their neighbors' effects. The condition of Theorem 3.2 requires that the norm of the vector $L\boldsymbol{\alpha} \in \mathbb{R}^n$ grows slower than $O(\sqrt{n})$. This condition is satisfied by a large set of $n-$dimensional vectors defined on many networks; the following proposition gives an example.

PROPOSITION 2. Assume the network is a $\sqrt{n}\times\sqrt{n}$ lattice. Then $\|L\boldsymbol{\alpha}\|^2 \leq n^c$ as long as $\boldsymbol{\alpha}$ is in the subspace spanned by $k$ smallest eigenvalues of $L$ for some $k \leq Cn^{\frac{1+c}{2}}$, where $C$ and $c$ are some constants and $c < 1$.

It is instructive to compare the MSE of our estimator with the MSE of the ordinary least squares (OLS) estimator, as well as the null model (which is what our estimator gives when the network has no edges). For OLS, we have

$$\hat{\boldsymbol{\beta}}_{OLS} = (X^TX)^{-1}X^T\boldsymbol{Y},\ \hat{\boldsymbol{\alpha}}_{OLS} = \bar{y}\mathbf{1},$$

where $\hat{\boldsymbol{\alpha}}_{OLS}$ is the common intercept. Compared to OLS, the RNC estimator reduces bias caused by the network-induced dependence among samples, and as a trade-off increases variance; thus intuitively, one would expect that the signal-to-noise ratio and the degree of cohesion in the network will determine which estimator performs better. From Theorem 3.1 and the basic properties of the OLS estimator (stated as Lemma 1 in the Appendix), it is easy to see that if

$$\left(\frac{n}{\nu} - 1\right)\sigma^2 \leq V(\boldsymbol{\alpha}) - \frac{\lambda^2}{\nu^2}\|L\boldsymbol{\alpha}\|^2 \tag{3.5}$$

where $V(\boldsymbol{\alpha}) = \sum_v(\boldsymbol{\alpha}_v - \bar{\boldsymbol{\alpha}})^2$, then the RNC estimator of the individual effects $\hat{\boldsymbol{\alpha}}$ has a lower MSE than that of $\hat{\boldsymbol{\alpha}}_{OLS}$. The left hand side of (3.5) represents the increase in variance induced by adding the individual effects, whereas the right hand size is the corresponding reduction in squared bias. When $\boldsymbol{\alpha}$ is sufficiently smooth over the network, $\|L\boldsymbol{\alpha}\|$ is negligible compared to other terms, and the condition essentially requires that the total variation of $\alpha_v$ around its average is larger than the total noise level. Similarly, for the coefficients $\beta$, if

$$\text{tr}((X^TX)^{-1})\frac{\sigma^2}{\nu} \le \|(X^TX)^{-1}X^T\boldsymbol{\alpha}\|^2 - \frac{\lambda^2}{\mu}\|L\boldsymbol{\alpha}\|^2 \tag{3.6}$$

then the RNC estimator $\hat{\boldsymbol{\beta}}$ has a lower MSE than $\hat{\boldsymbol{\beta}}_{OLS}$. Again, the two sides of the inequality represent the increase in variance and the reduction in squared bias, respectively. The null model gives an estimate for $\boldsymbol{\beta}$ identical to $\hat{\boldsymbol{\beta}}_{OLS}$, so the same comparison applies. The null model estimate of $\boldsymbol{\alpha}$ involves more terms and the corresponding tuning parameter and does not result in clear comparison. However, we demonstrate the difference numerically by the next example and by our simulation study in Section 4. The exact formula for the null model estimation error is given by Lemma 3 in Appendix A.

Example 1. We illustrate the bias-variance trade-off on a simple example. Suppose we have a network with $n = 300$ nodes which consists of three disconnected components $G_1$, $G_2$, $G_3$, of 100 nodes each. Each component is generated as an Erdos-Renyi graph, with each pair of nodes forming an edge independently with probability 0.05. Individual effects $\alpha_i$ are generated independently from $\mathcal{N}(\eta_{c_i}, 0.1^2)$, where $c_i \in \{1, 2, 3\}$ is the component to which nodes $i$ belongs, $\eta_1 = -1$, $\eta_2 = 0$, $\eta_3 = 1$. We set $\lambda = 0.1$. Substituting the expectation $EA$ for $A$, we have $\nu \approx 0.5$, $\|L\boldsymbol{\alpha}\|^2 \approx 105$, and $V(\boldsymbol{\alpha}) \approx 203$. Then as long as the noise variance $\sigma < 0.57$, (3.5) will be satisfied. Similarly, $X^TX \approx nI_2$, and $\|X^T\boldsymbol{\alpha}\|^2 \approx 406$ in expectation. Thus (3.6) holds and the RNC is beneficial if $\sigma < 0.54$ (approximately). The bias-variance trade-off in the mean squared prediction errors (MSPE) can be demonstrated explicitly when varying $\lambda$; Figure 1 shows this trade-off between bias and variance together with the OLS baseline when $\sigma = 0.5$. The MSPEs of OLS and the null model are also shown. Note that this calculation for RNC is based on conservative bounds. In reality the RNC is going to be beneficial for a larger range of $\sigma$ values.

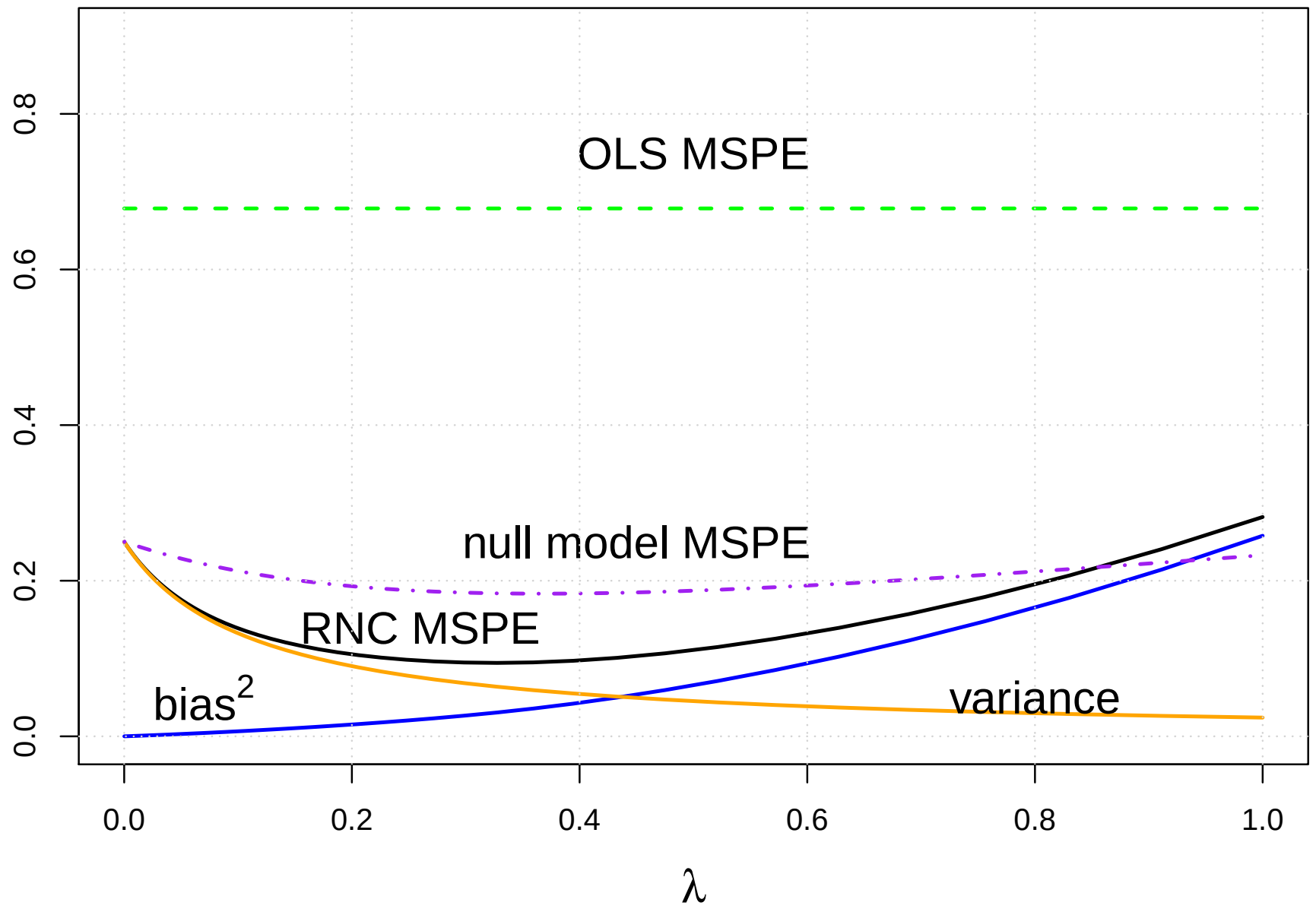

Fig 1: Mean squared prediction error $\mathbb{E}\|\hat{\boldsymbol{Y}} - \mathbb{E}\boldsymbol{Y}\|^2/n$ and the bias-variance trade-off of the RNC estimator (based on the upper bound (3.4) in Theorem 3.1), in the setting of Example 1 with $\sigma = 0.5$.

Remark 5. If we use (2.6) and are willing to make strong assumptions about the distribution as in the Bayesian interpretation, it can be shown (see Searle, Casella and McCulloch (2009), Ch. 7 for details) that $\hat{\boldsymbol{\alpha}}$ is the best linear unbiased predictor (BLUP) of $\boldsymbol{\alpha}$ and $\hat{\boldsymbol{\beta}}$ is the best linear unbiased estimator (BLUE) of $\beta$. However, these are strong assumptions which we prefer to avoid.

Finally, we investigate the effects of graph sparsification, proposed in Section 2.6 to reduce computational cost. For any $\epsilon > 0$, let $L^*$ be the Laplacian of a network on the same nodes satisfying

$$(1-\epsilon)L \preceq L^* \preceq (1+\epsilon)L. \tag{3.7}$$

In addition, let $\hat{\boldsymbol{\theta}}$ be the minimizer of

$$f(\boldsymbol{\theta}) = \ell(\boldsymbol{\alpha} + X\boldsymbol{\beta}; \boldsymbol{Y}) + \lambda \boldsymbol{\alpha}^T L \boldsymbol{\alpha}, \tag{3.8}$$

and $\hat{\boldsymbol{\theta}}^*$ be the minimizer of

$$f^*(\boldsymbol{\theta}) = \ell(\boldsymbol{\alpha} + X\boldsymbol{\beta}; \boldsymbol{Y}) + \lambda \boldsymbol{\alpha}^T L^* \boldsymbol{\alpha}, \tag{3.9}$$

where $\ell$ can be a general loss function, such as the sum of squared errors in linear model or the negative log-likelihood in GLM.

THEOREM 3.3. *Given two Laplacians $L$ and $L^*$ satisfying (3.7) for $0 < \epsilon < 1/2$, assume $\ell$ in (3.8) is twice differentiable and $f$ is strongly convex with $m > 0$, such that for any $\boldsymbol{\theta} = (\boldsymbol{\alpha}, \boldsymbol{\beta}) \in \mathbb{R}^{n+p}$,*

$$\bigtriangledown^2 f(\boldsymbol{\theta}) \succeq m I_{n+p}.$$

*Then $\hat{\boldsymbol{\theta}}$ and $\hat{\boldsymbol{\theta}}^*$ minimizing (3.8) and (3.9) respectively, with the same $\lambda$, satisfy*

$$\|\hat{\boldsymbol{\theta}}^* - \hat{\boldsymbol{\theta}}\|^2 \le \frac{2\epsilon\lambda}{m} \min\left(2\hat{\boldsymbol{\alpha}}^T L \hat{\boldsymbol{\alpha}} + |\hat{\boldsymbol{\alpha}}^T L \hat{\boldsymbol{\alpha}} - \hat{\boldsymbol{\alpha}}^{*T} L^* \hat{\boldsymbol{\alpha}}^*| + 2\epsilon \hat{\boldsymbol{\alpha}}^{*T} L^* \hat{\boldsymbol{\alpha}}^*\ , \ \frac{2\epsilon\lambda}{m} \lambda_1(L)^2 \|\hat{\boldsymbol{\alpha}}\|^2\right). \tag{3.10}$$

The proof is given in the Appendix. Theorem 3.3 can be seen as a generalization of the result of Sadhanala, Wang and Tibshirani (2016) for point estimation by Laplacian smoothing (or krigging) for Gaussian and binary data. Our bound is slightly better than that of Sadhanala, Wang and Tibshirani (2016).

REMARK 6. The term $\hat{\boldsymbol{\alpha}}^T L \hat{\boldsymbol{\alpha}}$ is the cohesion penalty and is expected to be small for estimated $\hat{\alpha}$. Further, we can expect both $|\hat{\boldsymbol{\alpha}}^T L \hat{\boldsymbol{\alpha}} - \hat{\boldsymbol{\alpha}}^{*T} L^* \hat{\boldsymbol{\alpha}}^*|$ and $\epsilon \hat{\boldsymbol{\alpha}}^{*T} L^* \hat{\boldsymbol{\alpha}}^*$ to be much smaller than $\hat{\boldsymbol{\alpha}}^T L \hat{\boldsymbol{\alpha}}$, and the first bound in (3.10) is typically much smaller than the second. Therefore, the bound is essentially

$$\|\hat{\boldsymbol{\theta}}^* - \hat{\boldsymbol{\theta}}\|^2 \lesssim \frac{4\epsilon\lambda}{m} \hat{\boldsymbol{\alpha}}^T L \hat{\boldsymbol{\alpha}}. \tag{3.11}$$

REMARK 7. The theorem shows that the squared error in estimation with an $\epsilon$-approximated Laplacian is decreasing linearly in $\epsilon$. In particular, it is easy to check that for the linear regression case, we have

$$\bigtriangledown^2 \ell(\boldsymbol{\theta}) = 2(\tilde{X}^T \tilde{X} + \lambda M).$$

Strong convexity always holds whenever RNC estimate exists, and the bound becomes

$$\|\hat{\boldsymbol{\theta}}^* - \hat{\boldsymbol{\theta}}\|^2 \lesssim \frac{2\epsilon\lambda \hat{\boldsymbol{\alpha}}^T L \hat{\boldsymbol{\alpha}}}{\lambda_n(\tilde{X}^T \tilde{X} + \lambda M)}. \tag{3.12}$$

Remark 8. Theorem 3.3 can also be viewed as a result on network misspecification. If the true network is observed with errors, but its Laplacian $L^*$ satisfies (3.7) and is close enough to the correct $L$, we have the same error bound for the estimate from the mispecified network. Another way to make the method more robust to errors in the network is to replace $L$ by a low-rank approximation to it, if we have reasons to believe a low-rank structure describes the underlying network well.

**4. Numerical performance evaluation.** In this section, we investigate the effects of including network cohesion on simulated data, in linear and logistic regression.

The simulated networks are generated from the stochastic block model with $n = 300$ nodes and $K = 3$ blocks. Under the stochastic block model, the nodes are assigned to blocks independently by sampling from a multinomial distribution with parameters $(\pi_1, \ldots, \pi_K)$. Then given block labels $c_i$ for $i = 1, \ldots, n$, the edges $A_{ij}$, $1 \le i < j \le n$, are generated as independent Bernoulli variables with $P(A_{ij} = 1) = B_{c_i c_j}$, where the $K \times K$ symmetric matrix $B$ contains probabilities of within-block and between-block connections. We set $\pi_1 = \pi_2 = \pi_3 = 1/3$, $B_{kk} = p_w = 0.2$, and $B_{kl} = p_b$ for all $k \ne l$. As in Example 1, the individual effects $\alpha_i$'s are generated independently from a normal distribution with the mean determined by the node's block label, $\mathcal{N}(\eta_{c_i}, s^2)$, where $\eta_1 = -1$, $\eta_2 = 0$, $\eta_3 = 1$, and the parameter $s$ controls how "cohesive" the $\alpha_i$'s within each block are. The predictor coefficients $\boldsymbol{\beta}$ are drawn independently from $\mathcal{N}(1, 1)$.

This simulation setting is not especially favorable to RNC since it does not satisfy the smoothness requirement of Theorem 3.2 except when $s = 0$. Moreover, because edges connecting different blocks give false information and edges within the same block are all exchangeable, an edge between two nodes does not give direct evidence of them having similar $\alpha$'s (except when $p_b = 0$). However, there is cohesion on the network in the sense that some $alpha$s are more similar to each other than to others, and we can vary the strength of cohesion by varying $s$; varying $p_b$ allows us to test robustness against "false" edges, meaning edges that do not indicate similarity.

We compare RNC to four other methods on these simulated networks: a baseline (OLS for continuous response and logistic regression for binary response), the null model, where the graph is empty and we simply add a ridge penalty on the individual effects, a fixed effects "oracle" model which knows the true blocks and uses the same $\alpha$ for all the nodes in the same block, and a mixed effects model which adds Gaussian random effects to the fixed effect model, fitting exactly the model that was used to generate the data. The tuning parameters are always selected by 10-fold cross-validation; however, the linear null model always makes the same out-of-sample predictions as OLS (Lemma 3 in the Appendix), for any value of $\lambda$, and thus cross-validation cannot be used to select the tuning parameter. This is a side effect of the bigger problem for the null model, which is its inability to make non-trivial out-of-sample predictions. Instead of cross-validation, we use the restricted maximum likelihood (REML) estimate under the corresponding linear mixed model framework for $\lambda = \sigma^2/\zeta^2$. The mixed effects model is also estimated by REML.

Four performance metrics are reported: the average mean squared error (MSE) of $\boldsymbol{\alpha}$ and $\boldsymbol{\beta}$, and in-sample and out-of-sample mean squared prediction errors (MSPEs). Figure 2 shows results as the variance parameter $s$ changes from 0 to 1 with $p_b = 0.02$. All methods get worse as $s$ increases and the signal-to-noise ratio goes down, as one would expect. The OLS is the worst on all measures since the other models incorporate the individual effects $\alpha$ and thus provide a better fit. However, incorporating $\alpha$ in the null model only helps with the in-sample error; for estimating $\beta$ and out-of-sample prediction, the null model is exactly the same as OLS. The RNC and the two oracle models generally perform much better and are fairly close to each other, with the oracle fixed effects model performing somewhat better on the in-sample error when $s$ is small and the oracle is close to the true model, and both the RNC and oracle mixed effects model outperforming the oracle fixed effects model for larger $s$ since they can adapt to the changing amount of cohesion over the network. Instead of using known blocks we could have also fitted them by one of the many available community detection methods, but that would only help if the underlying model does indeed have communities. The RNC, on the other hand, does not require an assumption of communities and can adapt to cohesion over many different types of underlying graphs.

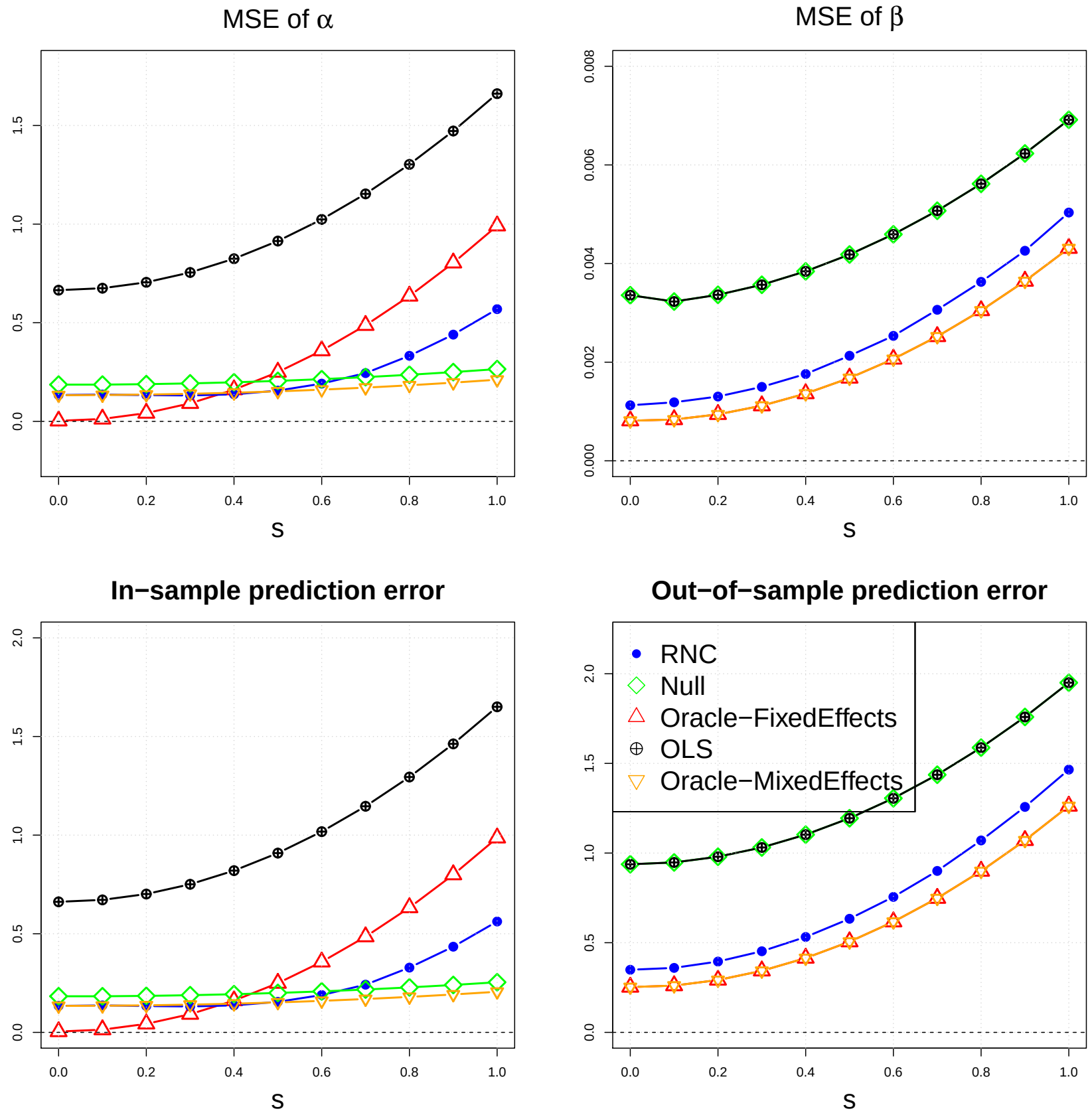

Fig 2: Linear regression with varying $s$ and $p_b = 0.02$. Performance is evaluated by the MSEs of $\boldsymbol{\alpha}$ and $\boldsymbol{\beta}$, and in-sample and out-of-sample mean squared prediction errors.

Figure 3 shows how the four performance metrics respond to an increase in $p_b$, the probability of "false" edges, with fixed $s = 0.1$. As expected, the performance of RNC degrades as $p_b$ increases. However, even when $p_b = 0.05$, when the ratio of within-block "true" edge probability to between-block "false" edge probability is only 4/3, RNC still does much better than OLS and the null model in estimating $\boldsymbol{\beta}$ and out-of-sample prediction.

Next, we use the same setting for generating the network, covariates, and parameters, but generate $\boldsymbol{Y}$ from the Bernoulli distribution with probabilities of success given by the logit function of $X^T\boldsymbol{\beta} + \boldsymbol{\alpha}$. We then estimate the parameters by fitting standard logistic regression and also logistic regression with our proposed network cohesion penalty. We fix a small value of the ridge regularization tuning parameter, $\gamma = 0.01$, as it is only added for numerical stability. We evaluate the methods by computing the average MSE of $\boldsymbol{\alpha}$, $\boldsymbol{\beta}$, and the vector of $n$ Bernoulli probabilities estimated as

$$\hat{p}_i = \frac{\exp(\boldsymbol{x}_i^T \hat{\boldsymbol{\beta}} + \hat{\alpha}_i)}{1 + \exp(\boldsymbol{x}_i^T \hat{\boldsymbol{\beta}} + \hat{\alpha}_i)},$$

as well as the probabilities on 50 hold-out samples. The latter two are analogues to in-sample and out-of-sample prediction errors in linear regression.

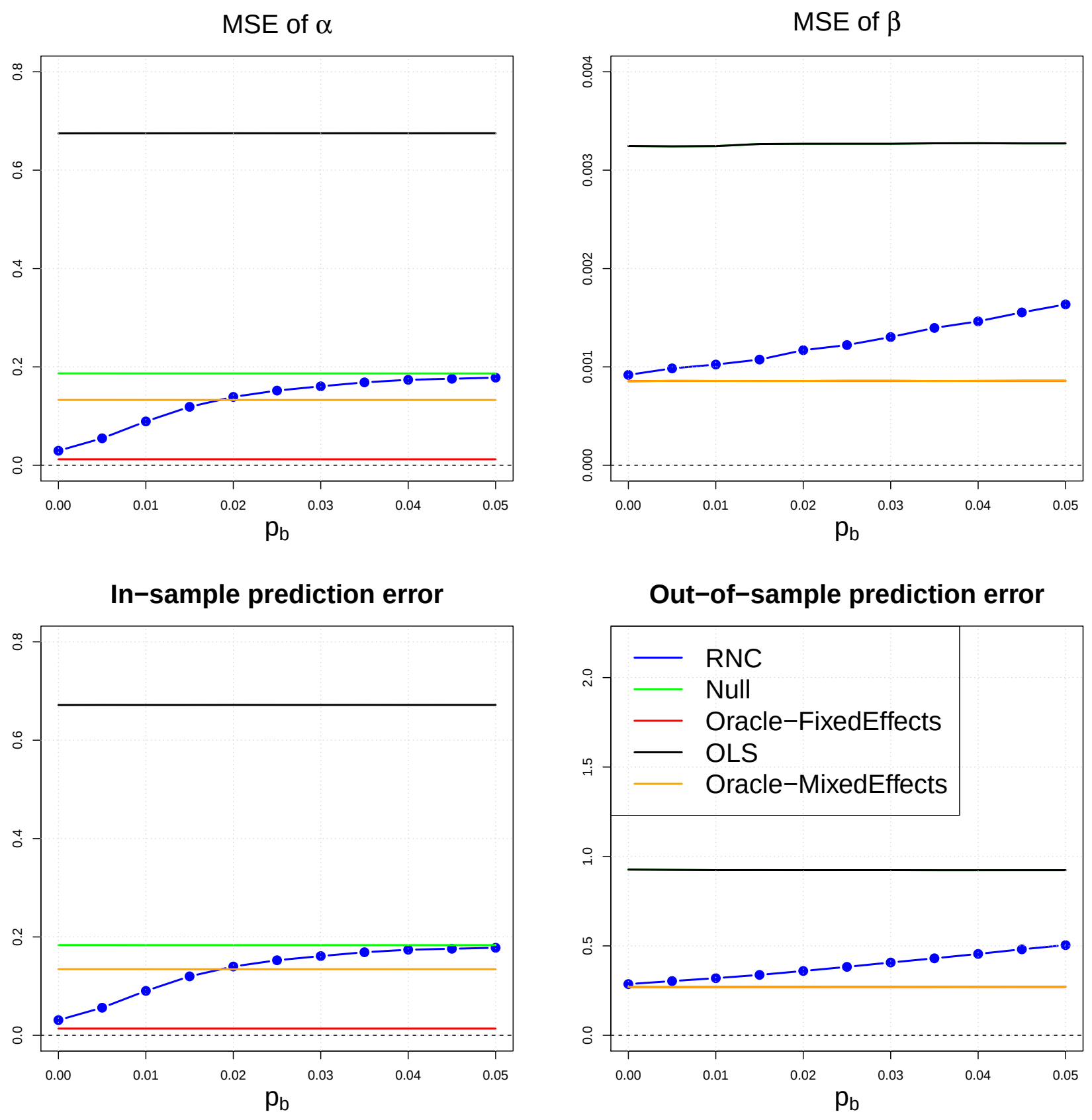

Fig 3: Linear regression with varying $p_b$ and $s = 0.1$. Performance is evaluated by the MSEs of $\boldsymbol{\alpha}$ and $\boldsymbol{\beta}$, and in-sample and out-of-sample mean squared prediction errors.

Figure 4 shows the average MSE of $\boldsymbol{\alpha}$, $\boldsymbol{\beta}$, and in-sample and out-of-sample probabilities as $s$ varies. The general pattern remains similar to linear regression. Although in this case the null model is no longer identical to regular logistic regression, it still gives nearly the same out-of-sample error. The oracle mixed effects model is the best, as it assumes the true model. In general, all methods deteriorate with increasing $s$, and while the logistic RNC does not perform quite as well as the oracle, it gets much closer to it than any other method. Figure 5 shows the metrics when varying $p_b$ from 0 to 0.05 with fixed $s = 0.1$. Again, the RNC outperforms regular logistic regression and the null model even when $p_b = 0.05$.

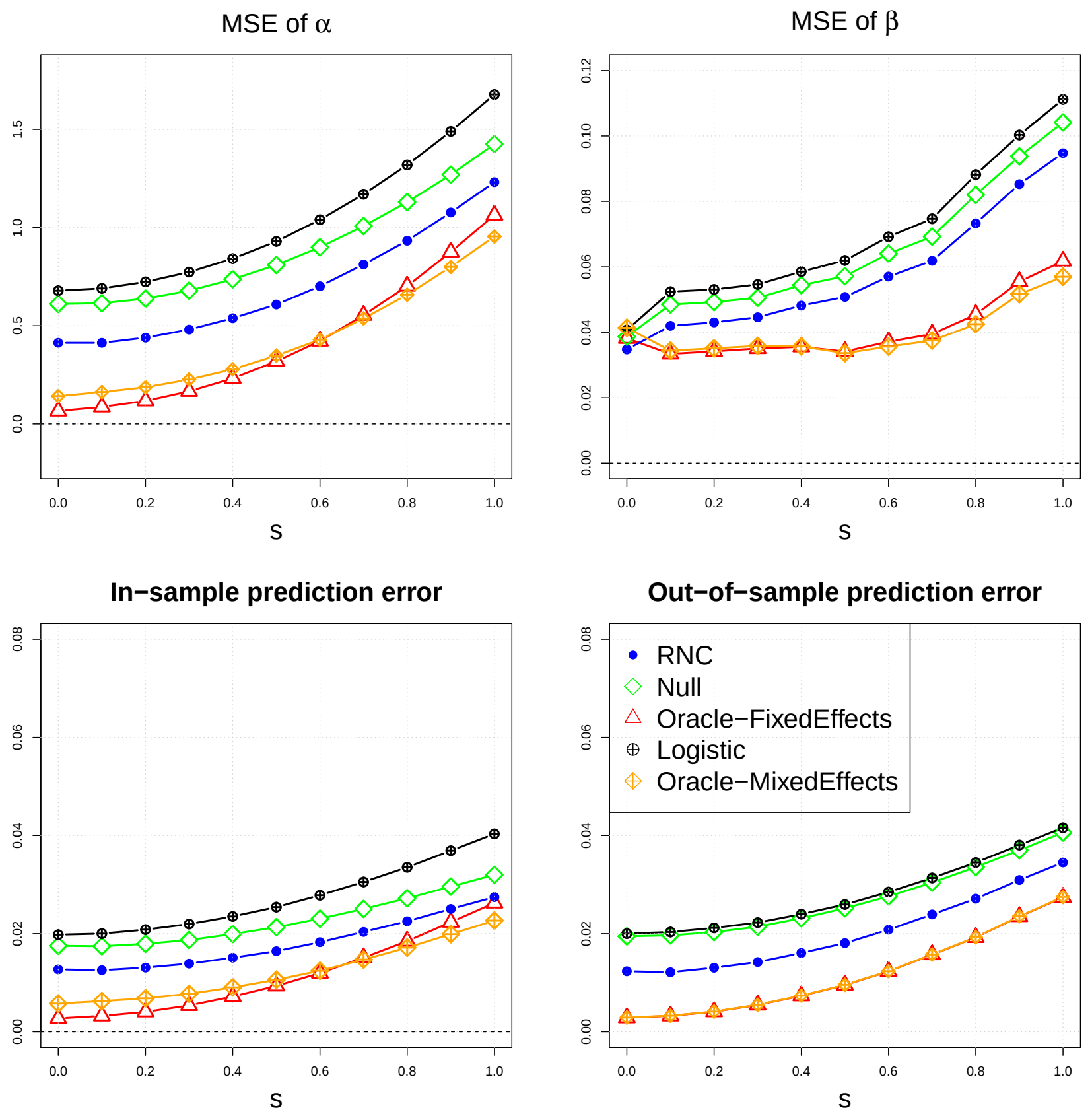

Fig 4: Performance logistic regression methods when varying $s$ and fixing $p_b = 0.02$, measured by the MSE of $\boldsymbol{\alpha}$, $\boldsymbol{\beta}$, in-sample and out-of-sample mean squared probability estimation errors.

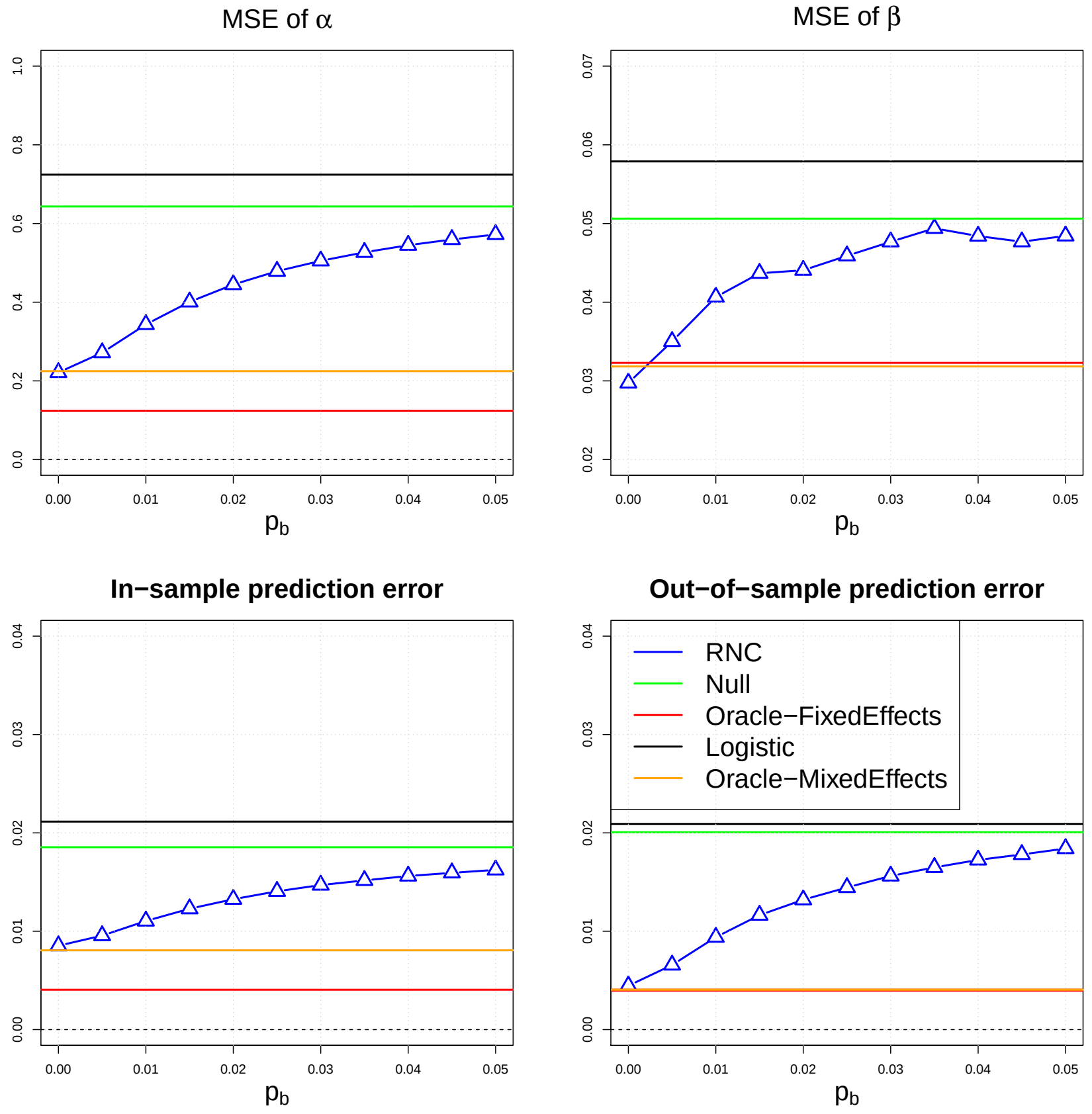

Fig 5: Performance of five logistic regression methods when varying $p_b$ and fixing $s = 0.1$, measured by the MSE of $\boldsymbol{\alpha}$, $\boldsymbol{\beta}$, in-sample and out-of-sample mean squared probability estimation errors.

We conclude this section with a simple example illustrating the graph sparsification approach to dense networks. We generate a weighted network with $n = 3000$ nodes, divided into three blocks of 1000 nodes each. All the within-block entries of the weighted adjacency matrix are 1 and the other entries are 0.1. Thus the network matrix is a fully dense matrix. The other settings are the same as in the linear regression simulation, and we compare the linear RNC estimator estimated using the original Laplacian $L$ to the one based on the sparsified $L^*$. Figure 6 shows the results as a function for different values of the approximation accuracy $\epsilon$'s, defined in (3.7). The top left plot shows the the sparsified matrix corresponding to $\epsilon = 0.15$, which has around 52% of all elements set to 0. The top right plot shows the observed approximation error $\|\hat{\boldsymbol{\theta}}^* - \hat{\boldsymbol{\theta}}\|^2$ and its theoretical upper bound (3.12). The theoretical bound is conservative but follows the same trend. Finally, the bottom plots of the difference in estimation errors for $\boldsymbol{\alpha}$ and $\boldsymbol{\beta}$ show that the difference between the sparsified and the original estimators goes to 0 as $\epsilon \to 0$, as it should, and that for moderate values of $\epsilon$ the differences are small and can go in either direction, which suggests an increase in variance but not much change in bias. Overall, in this example sparsification provides a reliable approximation to the original RNC estimator, and is a useful tool to save computational time for large dense networks.

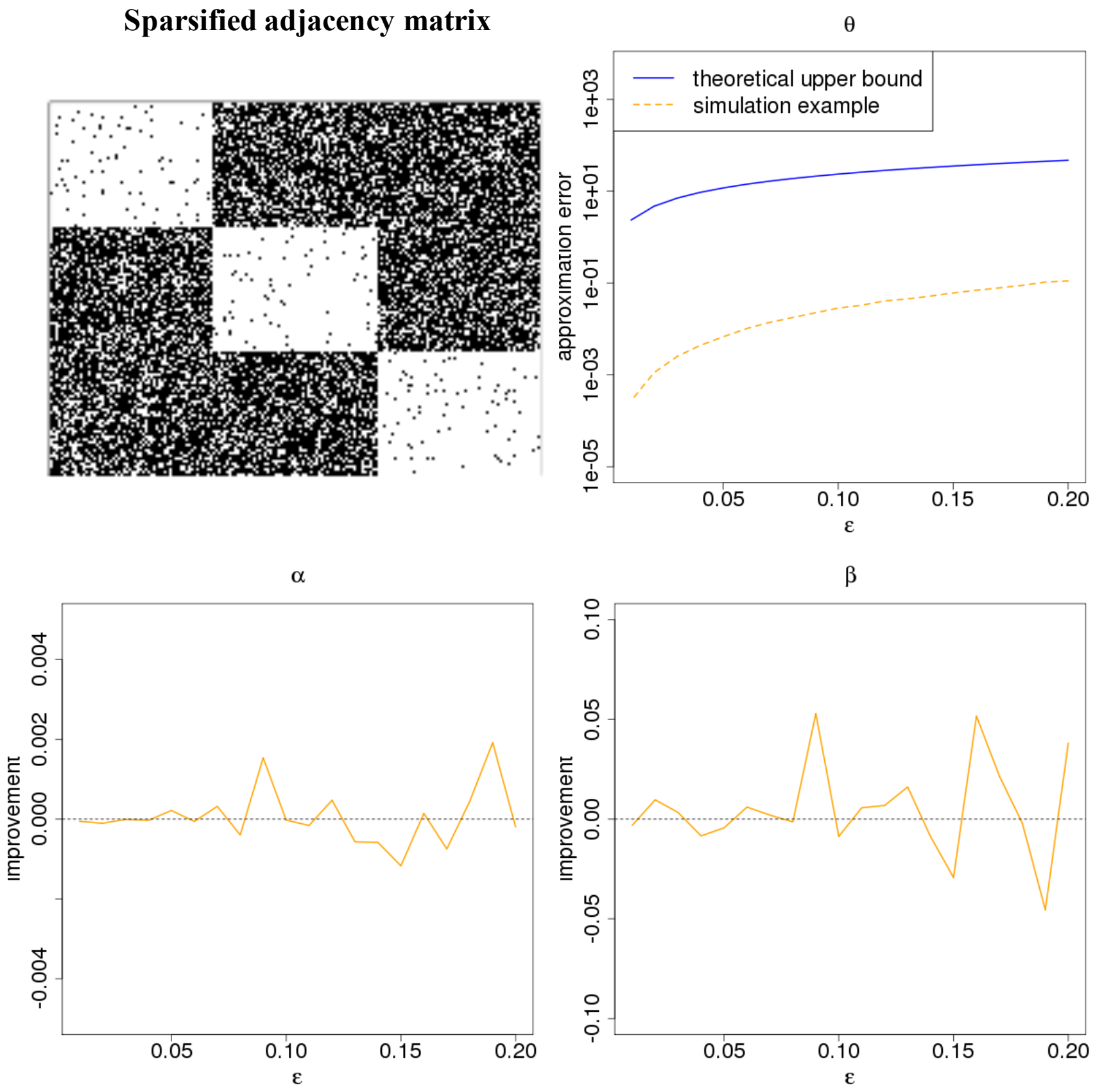

Fig 6: Top left: the adjacency matrix of the sparsified network for $\epsilon = 0.15$ (white indicates a nonzero entry, black is a zero entry); Top right: $\|\hat{\boldsymbol{\theta}}^* - \hat{\boldsymbol{\theta}}\|^2$ and the bound (3.12); Bottom left: relative improvement of the sparsified estimator $\boldsymbol{\alpha}^*$ over the original estimator $\hat{\boldsymbol{\alpha}}$, that is, $1 - \mathrm{MSE}_{\boldsymbol{\alpha}^*}/\mathrm{MSE}_{\hat{\boldsymbol{\alpha}}}$; Bottom right: relative improvement of the sparsified estimator $\boldsymbol{\beta}^*$ over the original estimator $\hat{\boldsymbol{\beta}}$.

**5. Analysis of the AddHealth Data.** We investigate the ability of our method to capture network effects and improve prediction in two applications using data from the AddHealth study (Harris, 2009). We will only use Wave I data in which both covariates and friendship networks are available. Our first test task is predicting students' recreational activity from their demographic covariates and their friendship networks; this was done via a network autoregressive model in Bramoullé, Djebbari and Fortin (2009), who used the in-school survey data. In order to be able to compare with their results directly, we also use the in-school data only for this task. The students were asked about friends at both in-school and in-home interviews, and the resulting networks are somewhat different. Our second application is predicting the age of first marijuana use, and the data on marijuana use are only available from the in-home interviews; thus for the second task we use the friendship network constructed from the in-home interviews. Prediction performance on these two tasks is presented in this section. Additional results on sensitivity to missing data are presented in Appendix D.

5.1. *Predicting recreational activity in adolescents: a linear model example.* This exact task on the AddHealth data was considered by Bramoullé, Djebbari and Fortin (2009), who incorporated peer effects into ordinary linear regression in via the auto-regressive model

$$y_v = \frac{1}{|N(v)|} \sum_{u \in N(v)} (\gamma y_u + \boldsymbol{x}_u^T \boldsymbol{\tau}) + \boldsymbol{x}_v^T \boldsymbol{\beta} + \epsilon_v, v \in V \ , \tag{5.1}$$

or, equivalently, in matrix form

$$Y = (I - \gamma D^{-1} A)^{-1} (D^{-1} A X \boldsymbol{\tau} + X \boldsymbol{\beta} + \boldsymbol{\epsilon}). \tag{5.2}$$

The authors called this the social interaction model (SIM), also sometimes called a "linear-in-means" model. In econometric terminology, the local average of responses models endogenous effects, and the local averages of predictors are the exogenous effects. This model generally requires multiple additional assumptions to be identifiable and distributionally compatible across different equations, an issue not considered by Bramoullé, Djebbari and Fortin (2009). It also loses the interpretation of predictor coefficients as the change in the predicted value corresponding to a unit increase in one predictor with all others fixed. When there are known groups in the data, fixed effects can be added to this model (Lee, 2007). In Bramoullé, Djebbari and Fortin (2009), SIM was applied to the AddHealth data to predict levels of recreational activity from a number of demographic covariates as well as the friendship network. The covariates used are age, grade, sex, race, whether born in the U.S. living with the mother, living with the father, mother's education, father's education, and parents' participation in the labor market. For some of the categorical variables, some of the levels were merged; refer to Bramoullé, Djebbari and Fortin (2009) for details. Recreational activity was measured by the number of clubs or organizations to which the student belongs, with "4 or more" recorded as 4. The histogram as well as the mean and standard deviation of recreational activity are shown in Figure 7. We used exactly the same variables with the same level merging.

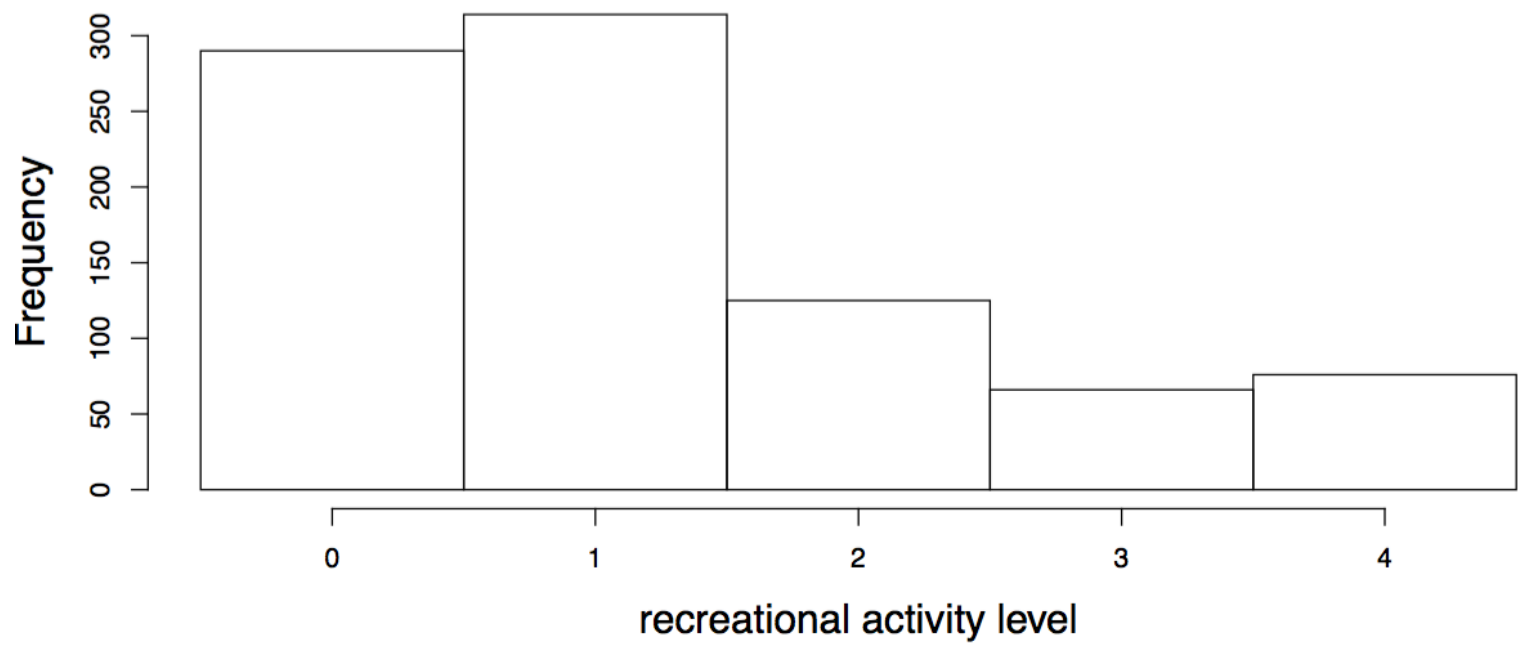

Fig 7: Histogram of the response, recreational activity level, from the data set used in the linear regression example. The mean recreational activity is 1.22, with standard deviation 1.23.

We compare performance of our proposed RNC method with the SIM model (5.1) from Bramoullé, Djebbari and Fortin (2009), and to regular linear regression without network effects implemented by ordinary least squares (OLS), with the same response and predictors as inBramoullé, Djebbari and Fortin (2009). As discussed, the null model always gives out-of-sample predictions identical to OLS, so we do not distinguish between them in this example. We also include boosting (with tree stumps as base learners) using the same covariates as OLS, which gives a measure of how much improvement relative to OLS can be achieved using the same covariates without a linearity assumption. Boosting was tuned by cross-validation using the R package of Hothorn et al. (2018). As an additional comparison with SIM, we also fit RNC with local averages of predictors as additional variables in the model (RNC-LA). We also apply the Bayesian model from Section 2.4, which is equivalent to the CAR model as discussed in Section 2.7. However, such a model can only make out-of-sample predictions if the entire network, including that of the test data, is available before training. Therefore, we implement this model as an oracle, including the entire network connecting the test and the training data at the training stage. We call this method "oracle-Bayes" to indicate it is using oracle information that is not available to all the other methods, and thus is not a fair competitor. The Bayesian estimates are computed as posterior medians from MCMC samples using the implementation of Lee (2013).identical

We use the largest school in the dataset, with 2350 students. For 1223 records with some missing values, we implemented conditional imputation, using random forests trained on all the variables without missing values. In Appendix D, we include a sensitivity analysis to the proportion of missing data, showing that our analysis is very robust. To order predictors, we randomly split the network in two connected subgraphs with similar sizes. We use one of these connected networks, with 898 data points, to perform variable selection, and the other network with 940 points for evaluating the models. The remaining 512 samples are not connected to either of the two networks and mostly consist of isolated nodes or isolated pairs; we remove them from the analysis since those are not going to be able to demonstrate peer effects.

We perform forward variable selection on the variable selection set, and then add variables in the selected order to the model fitted on the other dataset. Doing variable selection and model evaluation on two separate data sets avoids introducing model selection bias into our estimated prediction error. The forward selection procedure starts with fitting an RNC model without any covariates, obtaining an estimate of $\hat{\boldsymbol{\alpha}}$ from this model, and then running standard forward selection adding one variable at a time to $\hat{\boldsymbol{\alpha}}$ which always remains in the model with a fixed coefficient 1. This ensures that selected variables are not acting as proxies to peer effects but are adding as much new information as possible.

To evaluate predictive performance, we randomly hold out 90 students (about 10%) from the model evaluation dataset as test data, and fit all the models on the rest. The variables are added to all the models one at a time in the order determined by the variable selection procedure. The procedure is repeated for 50 independent random data splits into training and test sets. The root mean squared errors (RMSEs) over these 50 splits are shown in Table 1. In each row, we report the results from a paired t-test over the 50 random splits for each model compared with RNC. It is clear that both SIM and RNC are able to improve predictions by using information from the network, but RNC is more effective at this in all models. Including local averages of predictors does not help RNC at all, indicating that the network effects it picks are distinct from and perhaps more informative than the ones reflected in local average. The oracle Bayes method does not perform as well as RNC either, though it uses more information. A potential explanation for this may be that the specific distribution assumptions that the Bayesian model imposes are not satisfied for this dataset; in particular, it might be a stretch to model the 4-level ordinal recreation activity variable as Gaussian. Using boosting does not help either, because the predictors are not particularly informative.

The network information is relatively more helpful: the RNC error using only network cohesion and none of the predictors is lower than the error of *any* model fitted by either OLS, SIM or oracle Bayes. As with any other prediction task, adding unhelpful covariates tends to corrupt performance, and RNC achieves the best performance with only one predictor in the model (father's education). Finally, the coefficients from both OLS and RNC regressions are reported in Table 4 of Appendix C. They are generally similar, suggesting that the network cohesion penalty does not fundamentally change interpretation of the coefficients, but improves prediction.

TABLE 1

*Root mean squared errors for predicting recreational activity, over 50 independent data splits into test (90 samples) and training sets. All methods are compared to RNC by a paired two-sample t-test, where ** indicates $p \leq 10^{-4}$ and * indicates $10^{-4} < p < 10^{-2}$. Each row adds the variable listed to the model in the previous row, in the order determined on a separate set by forward selection with network cohesion effects included.*

| model | OLS & Null | Boosting | SIM | RNC | RNC-LA | oracle-Bayes |
|---|---|---|---|---|---|---|
| no covariates | 1.217 ** | 1.214 ** | 1.177 ** | 1.157 | 1.157 | 1.165 * |
| + father's education | 1.215 ** | 1.210 ** | 1.180 ** | 1.156 | 1.160 * | 1.165 |
| + race | 1.213 ** | 1.212 ** | 1.178 ** | 1.158 | 1.164 * | 1.163 |
| + age | 1.214 ** | 1.212 ** | 1.177 ** | 1.158 | 1.163 | 1.161 |
| + mother's education ** | 1.216 ** | 1.213 ** | 1.179 ** | 1.160 | 1.167 | 1.171 |
| + born in the US | 1.217 ** | 1.214 ** | 1.179 ** | 1.161 | 1.169 | 1.168 |
| + sex | 1.211 ** | 1.208 ** | 1.174 * | 1.157 | 1.161 | 1.167 |
| + parents in labor market | 1.214 ** | 1.209 ** | 1.179 ** | 1.159 | 1.165 | 1.172 * |
| + living with mother | 1.216 ** | 1.210 ** | 1.182 ** | 1.161 | 1.169 | 1.174 * |
| + living with father | 1.218 ** | 1.210 ** | 1.186 ** | 1.163 | 1.174 * | 1.172 |
| + grade | 1.219 ** | 1.210 ** | 1.188 ** | 1.163 | 1.176 * | 1.175 * |

REMARK 9. For fair comparison with Bramoullé, Djebbari and Fortin (2009), we formulate the problem of predicting recreational activity level as a linear regression problem. However, given the ordinal nature of the response, an alternative option may be using ordinal regression with network cohesion.

5.2. *Predicting the risk of adolescent marijuana use.* While many prediction tasks can be addressed with linear or logistic regression, there are settings where survival analysis is the only appropriate tool. In the AddHealth survey, the students were asked "How old were you when you tried marijuana for the first time?", and the answer can either be age (an integer up to 18) or "never", which is a censored observation of age of first use. A survival model is thus the appropriate prediction tool. Here we apply Cox's proportional hazard model, with network cohesion, to the largest community in the dataset with 1862 students from the Wave I in-home interview (this question was only asked in the in-home interviews). The friendship network is also based on in-home data for consistency; there are 2820 additional covariates on each student collected from the in-home surveys.

As before, missing values are imputed with conditional imputation using random forests, with covariates without missing values as predictors. However, we deleted variables that had missing values due to questionnaire design, and variables with more than half the values missing. This left us with 218 variables in total (since there are so many variables in the in-home survey, there are many missing values). As in the previous example, we split the data randomly into two connected components of roughly equal sizes, 645 observations for variable selection and 647 observations for model evaluation. The remaining isolated nodes or pairs are removed. The variable selection step is implemented as in the previous example, with network cohesion effects in the model. Five strongest predictors are selected, with the additional requirement that each survey category (survey questions were grouped) has no more than one variable selected. We then use a regular forward selection algorithm to determine the order in which these five variables should be added to the model.

Given the selected variables and the order in which to add them, we fit the regular Cox's model, the null model, boosting variant of the Cox's model Bühlmann and Hothorn (2007) and the RNC for survival on the model evaluation data set. The null model is numerically nearly identical to the regular Cox's model. We also include a naive extension of the social interaction model (SIM) (5.1) to survival analysis, including the neighborhood averages of $x$'s as extra covariates. However, the neighborhood averages of $y$'s cannot be computed here, since many of the $y$'s are censored and it is not clear how to extend the autoregressive component of the model to survival data. We also include "RNC-LA" again, which adds all the local averages of predictors to the RNC model. In the survival model, RNC can be fitted with no covariates, but this is not possible for the regular Cox's model or SIM since partial likelihood is not defined without covariates.

Evaluating predictive performance of survival models is not straightforward; we use the survival ROC curve suggested by Song and Zhou (2008). We calculate the prediction ROC curve for each age between 14 and 17 (most age values fall into this range), then integrate the area under curve (AUC) over age to get a measure of overall prediction performance. We randomly select 60 nodes (about 10%) as the test set and use the remaining nodes and their induced sub-network as the training set. This is independently repeated 50 times and the average integrated AUC (iAUC) over the 50 replications is used to evaluate performance. For simplicity of comparisons, we fixed the tuning parameter $\lambda = 0.005$ for all models, based on validation on a different school, and set $\gamma = 0.1$. This results in a conservative comparison of our method to Cox's model, since tuning each RNC fit separately can only improve its performance.

TABLE 2

*Average integrated AUC (iAUC) for survival prediction ROC curves for age 14-17, over 50 random splits of the data into training and test sets. All methods are compared with RNC by a paired two-sample t-test. ** indicates $p \leq 10^{-4}$ and * indicates $10^{-4} < p < 10^{-2}$. Each row adds the variable listed to the model in the previous row, in the order determined on a separate set by forward selection with network cohesion effects included.*

| model | Cox & Null | Boosting | SIM | RNC | RNC-LA |
|---|---|---|---|---|---|
| no covariates | – | – | – | 0.606 | 0.606 |
| + ever tried cigarette smoking | 0.657 ** | 0.668 ** | 0.663 ** | 0.709 | 0.703 ** |
| + deliberately damaged others' property | 0.700 ** | 0.714 ** | 0.707 ** | 0.735 | 0.736 |
| + times of being jumped in past 12 months | 0.713 ** | 0.722 ** | 0.733 * | 0.740 | 0.758 ** |
| + how often to wear seatbelt in a car | 0.721 ** | 0.729 ** | 0.743 | 0.745 | 0.765 ** |
| + received school suspension | 0.727 ** | 0.742 * | 0.743 | 0.748 | 0.766 ** |

Table 2 shows the average iAUC results. All models improve or stay the same with additional predictors. All methods that use the network information always do better than the regular Cox's model with the same covariates. RNC always outperforms SIM, and RNC-LA improves upon RNC for models with more covariates, but not for the smaller ones. This may suggest that some predictors' local averages are more helpful than others; however, including any local averages distorts the meaning of the coefficients. Overall, the network cohesion effect in predicting marijuana usage is clearly useful. Boosting is also able to improve the predictive power over the regular Cox's model; however, RNC outperforms boosting in all cases. This suggests that the network effect is more important than nonlinear covariate effects (as modeled by boosting), though boosting is already much better than models with only linear effects.

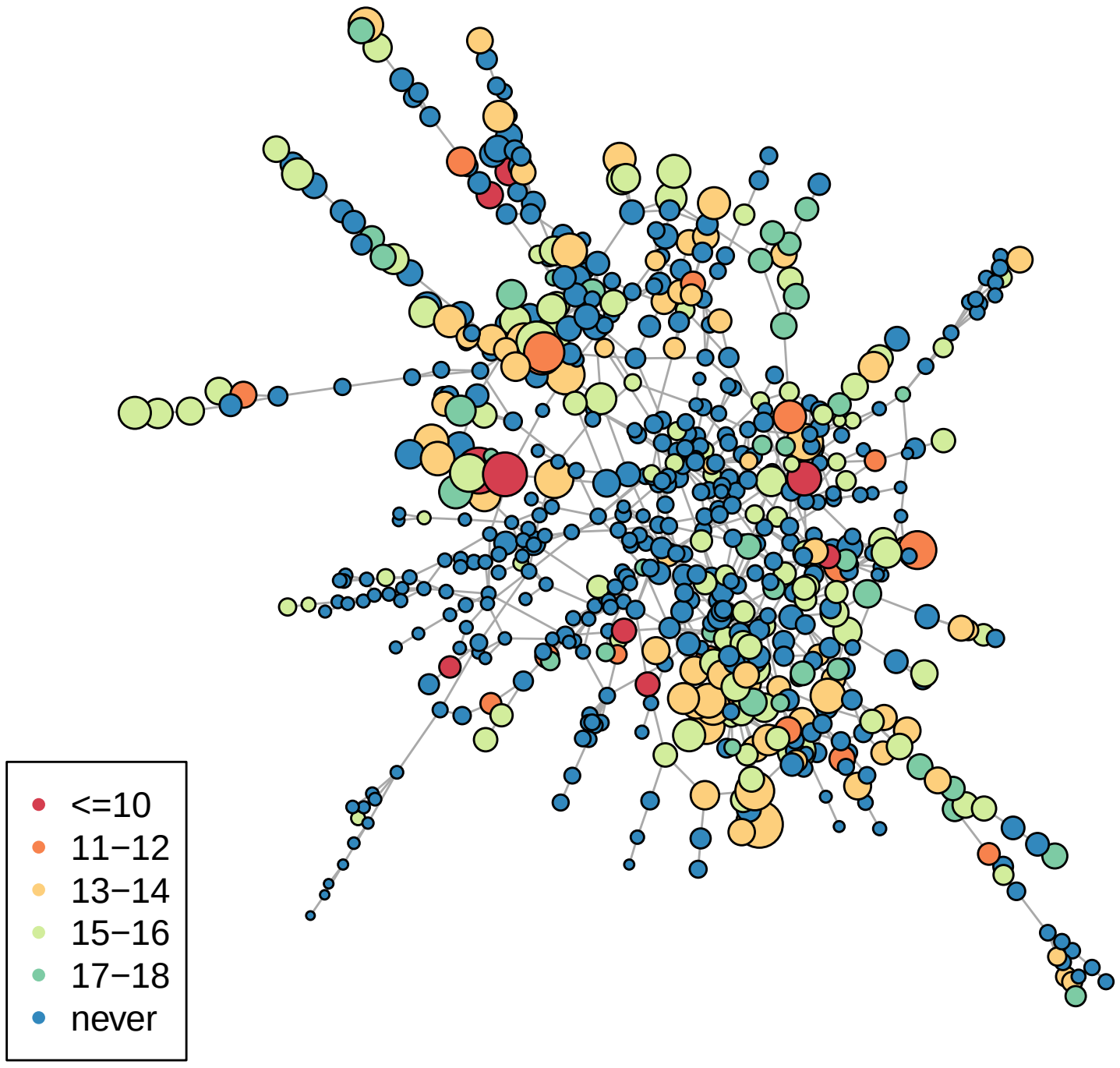

Fig 8: Age of first marijuana risk use shown on the friendship network. Node size represents the individual's hazard, and node color represents the observed age of first use.

The estimated individual hazards $\exp(\hat{\alpha}_v)$'s are shown in Figure 8, represented by node size, together with the friendship network and the observed age represented by node color. The cohesion effect can be seen in both the data itself and in the estimated hazards.

Table 3 shows the coefficients of the regular Cox's model and the RNC model. They are overall similar, though it appears that for most variables the coefficient is slightly reduced with the addition of network effects. This makes sense since some of the covariates are also likely cohesive Michell and West (1996); Pearson and Michell (2000); Pearson and West (2003) and can serve as proxies to peer effects, thus appearing to be more influential than they really are by themselves.

TABLE 3
*Estimated covariate coefficients from regular Cox's model and RNC for first age of marijuana use prediction.*

| covariate | Cox & Null | RNC | p-value (from Cox) |
|---|---|---|---|
| ever tried cigarette smoking | 1.627 | 1.370 | $< 10^{-6}$ |
| deliberately damaged others' property | 0.348 | 0.367 | $< 10^{-4}$ |
| times of being jumped in past 12 months | -0.122 | -0.191 | 0.077 |
| how often wears seatbelt | 0.288 | 0.283 | 0.007 |
| received school suspension | 0.633 | 0.473 | $< 10^{-6}$ |

**6. Discussion.** We have proposed a general computationally efficient framework for introducing network cohesion effects into prediction problems, without losing the interpretability and meaning of the original predictors. For the regression setting, we also derived conditions for when this approach outperforms regular regression and have shown the proposed estimator is consistent. In general, we can view RNC as another example of benefits of regularization when there are more parameters than one can estimate with the data available. Encouraging network cohesion implicitly reduces the number of free parameters, somewhat in the same spirit as the fused lasso penalty (Tibshirani et al., 2005). There are important differences, however; we have a computationally efficient way to use the available network data whereas the fused lasso optimization problem is hard to solve, and we can explicitly assess the trade-off in bias and variance that results from encouraging cohesion.

A future direction to explore is understanding the behavior of network cohesion on different kinds of networks. The large literature on random graph models for networks gives many options for modeling the network as random rather than treating it as fixed, as we did here; we would expect that some types of networks spread cohesion over the network faster than others. While we focused on prediction in this paper, the cohesion penalty may also turn out to be useful in causal inference on networks when such inference is possible. Formal inference under cohesion, such as confidence intervals and hypothesis tests, are also left for future work.

**Acknowledgements.** We thank the Associate Editor and two referees for many helpful suggestions that greatly improved the paper. This research was partly supported by ONR grant N000141612910, NSF grant DMS-1159005, DMS-1407698 and NIH grant R01GM096194. This research uses data from Add Health, a program project designed by J. Richard Udry, Peter S. Bearman, and Kathleen Mullan Harris, and funded by a grant P01-HD31921 from the Eunice Kennedy Shriver National Institute of Child Health and Human Development, with cooperative funding from 17 other agencies. Special acknowledgment is due Ronald R. Rindfuss and Barbara Entwisle for assistance in the original design. Persons interested in obtaining Data Files from Add Health should contact Add Health, The University of North Carolina at Chapel Hill, Carolina Population Center, 206 W. Franklin St., Chapel Hill, NC 27516-2524 (addhealth_contractsunc.edu). No direct support was received from grant P01-HD31921 for this analysis.

## APPENDIX A: PROOFS

Proof of Proposition 1. The first claim follows directly from the fact that $\mathbf{1}$ is an eigenvector of $P_{X^\perp} + \lambda L$ with eigenvalue 1, since $\mathbf{1} \in \mathrm{col}(X)^\perp$ and $L\mathbf{1} = 0$. To show the second claim, note that the minimum eigenvalue of $P_{X^\perp} + \lambda L$ is the solution of the optimization problem

$$\min_{\|\boldsymbol{u}\|=1} \boldsymbol{u}^T(P_{X^\perp} + \lambda L)\boldsymbol{u}.$$

Assume $\boldsymbol{u} = \boldsymbol{u}_1 + \boldsymbol{u}_2$, where $\boldsymbol{u}_1 \in \mathrm{col}(X)^\perp$, $\boldsymbol{u}_2 \in \mathrm{col}(X)$ and $\|\boldsymbol{u}_1\|^2 + \|\boldsymbol{u}_2\|^2 = 1$. Then the objective function can be rewritten as

$$\lambda \boldsymbol{u}^T L \boldsymbol{u} + \|\boldsymbol{u}_1\|^2.$$

This is zero if and only if $\|\boldsymbol{u}_1\| = 0$ and $\boldsymbol{u}^T L \boldsymbol{u} = 0$, but these two contradict Assumption 1. As discussed in Section 2.2, the RNC estimator exists whenever $P_{X^\perp} + \lambda L$ is invertible, which shows that the RNC estimate exists. □

One formula that will be used frequently later is the decomposition of MSE for a vector estimation:

$$\mathbb{E}\|\hat{\boldsymbol{\theta}} - \boldsymbol{\theta}\|^2 = \|\mathbb{E}\hat{\boldsymbol{\theta}} - \boldsymbol{\theta}\|^2 + \mathrm{tr}(\mathrm{Var}(\hat{\boldsymbol{\theta}})),$$

in which we call the second term total variance of $\hat{\boldsymbol{\theta}}$.

We first derive the bias and variance of both the OLS and the RNC estimators. We use $b(\cdot)$ to denote the bias of an estimator. The bias, variance and MSE of the OLS estimator are standard. We state the MSE here for completeness without proof.

Lemma 1. For the OLS estimator given by

$$\hat{\boldsymbol{\beta}}_{OLS} = (X^TX)^{-1}X^T\boldsymbol{Y},\ \hat{\boldsymbol{\alpha}}_{OLS} = \bar{y}\mathbf{1},$$

we have

$$\mathrm{MSE}(\hat{\boldsymbol{\alpha}}_{OLS}) = \|\bar{\boldsymbol{\alpha}}\mathbf{1} - \boldsymbol{\alpha}\|^2 + \frac{\sigma^2}{n},$$
$$\mathrm{MSE}(\hat{\boldsymbol{\beta}}_{OLS}) = \|(X^TX)^{-1}X^T\boldsymbol{\alpha}\|^2 + \sigma^2\mathrm{tr}((X^TX)^{-1}),$$

$$\mathbb{E}\|\hat{\boldsymbol{Y}}_{OLS} - \mathbb{E}\boldsymbol{Y}\|^2 = \|(\frac{1}{n}\mathbf{1}\mathbf{1}^T + X(X^TX)^{-1}X^T)\boldsymbol{\alpha} - \boldsymbol{\alpha}\|^2 + \sigma^2\|\frac{1}{n}\mathbf{1}\mathbf{1}^T + X(X^TX)^{-1}X^T\|_F^2.$$

LEMMA 2. The bias of the RNC estimator is given by

$$b(\hat{\boldsymbol{\theta}}) = -\lambda(\tilde{X}^T\tilde{X} + \lambda M)^{-1}M\boldsymbol{\theta}. \tag{A.1}$$

Equivalently, one can write it in the following decomposed form:

$$b(\hat{\boldsymbol{\theta}}) = (b(\hat{\boldsymbol{\alpha}})^T, ((X^TX)^{-1}X^Tb(\hat{\boldsymbol{\alpha}}))^T)^T, \tag{A.2}$$

where $b(\hat{\boldsymbol{\alpha}}) = -(\frac{1}{\lambda}P_{X^\perp} + L)^{-1}L\boldsymbol{\alpha}$, and $P_X = X(X^TX)^{-1}X^T$ is the projection matrix onto $\mathrm{col}(X)$.

The variance of the RNC estimator is given by

$$\mathrm{Var}(\hat{\boldsymbol{\theta}}) = \sigma^2(\tilde{X}^T\tilde{X} + \lambda M)^{-1}\tilde{X}^T\tilde{X}(\tilde{X}^T\tilde{X} + \lambda M)^{-1} \preceq \sigma^2(\tilde{X}^T\tilde{X} + \lambda M)^{-1}.$$

PROOF. For the bias term,

$$\begin{aligned} b(\hat{\boldsymbol{\theta}}) &= \mathbb{E}(\tilde{X}^T\tilde{X} + \lambda M)^{-1}\tilde{X}^TY - \boldsymbol{\theta} \\ &= (\tilde{X}^T\tilde{X} + \lambda M)^{-1}\tilde{X}^T\tilde{X}\boldsymbol{\theta} - \boldsymbol{\theta} \\ &= -\lambda(\tilde{X}^T\tilde{X} + \lambda M)^{-1}M\boldsymbol{\theta}. \end{aligned}$$

Note that we have $M\boldsymbol{\theta} = \begin{bmatrix} L\boldsymbol{\alpha} \\ 0 \end{bmatrix}$. By the block matrix inverse formula, we have

$$(\tilde{X}^T\tilde{X} + \lambda M)^{-1} =$$

$$\begin{bmatrix} (P_{X^\perp} + \lambda L)^{-1} & (P_{X^\perp} + \lambda L)^{-1}X(X^TX)^{-1} \\ (X^TX)^{-1}X^T(P_{X^\perp} + \lambda L)^{-1} & (X^TX)^{-1} + (X^TX)^{-1}X^T(P_{X^\perp} + \lambda L)^{-1}X(X^TX)^{-1} \end{bmatrix}.$$

Then (A.2) follows directly from decomposing the bias vector into the $\boldsymbol{\alpha}$ and $\boldsymbol{\beta}$ parts.

The variance can be calculated by the standard OLS formula taking $\tilde{X}$ as the design matrix. The positive semi-definiteness follows from the fact that

$$X^TX \preceq X^TX + \lambda M$$

whenever $M$ is positive semi-definite. □

From Lemma 2 and the bias-variance decomposition, we can directly get the closed form expressions for the MSE of RNC estimation. In particular,

$$\begin{aligned} \mathrm{MSE}(\boldsymbol{\theta}) =& \|\lambda(P_{X^\perp} + \lambda L)^{-1}L\boldsymbol{\alpha}\|^2 + \|\lambda(X^TX)^{-1}X^T(P_{X^\perp} + \lambda L)^{-1}L\boldsymbol{\alpha}\|^2 \\ &+ \sigma^2\mathrm{tr}((\tilde{X}^T\tilde{X} + \lambda M)^{-1}\tilde{X}^T\tilde{X}(\tilde{X}^T\tilde{X} + \lambda M)^{-1}). \end{aligned} \tag{A.3}$$

PROOF OF THEOREM 3.1. Note that $P_{X^\perp} + \lambda L \succeq \nu I$. Thus the squared bias term for $\boldsymbol{\alpha}$ is

$$\|\lambda(P_{X^\perp} + \lambda L)^{-1} L\boldsymbol{\alpha}\|^2 \le \frac{\lambda^2}{\nu^2}\|L\boldsymbol{\alpha}\|^2.$$

The total variance of $\hat{\boldsymbol{\alpha}}$ can be upper bounded by

$$\mathrm{tr}(\sigma^2(P_{X^\perp} + \lambda L)^{-1}) \le \frac{\sigma^2}{\nu}\mathrm{tr}(I) = \frac{n\sigma^2}{\nu}.$$

Thus the bound (3.2) on MSE($\hat{\boldsymbol{\alpha}}$) follows.

From Lemma 2, we have

$$\begin{aligned}
\|b(\hat{\boldsymbol{\beta}})\|^2 =& b(\hat{\boldsymbol{\alpha}})^T X(X^TX)^{-1}(X^TX)^{-1}X^T b(\hat{\boldsymbol{\alpha}}) \\
\le& \frac{1}{\mu} b(\hat{\boldsymbol{\alpha}})^T X(X^TX)^{-1}(X^TX)(X^TX)^{-1}X^T b(\hat{\boldsymbol{\alpha}}) \\
=& \frac{1}{\mu} b(\hat{\boldsymbol{\alpha}})^T X(X^TX)^{-1}X^T b(\hat{\boldsymbol{\alpha}}) = \frac{1}{\mu} b(\hat{\boldsymbol{\alpha}})^T(P_X b(\hat{\boldsymbol{\alpha}})) \\
=& \frac{1}{\mu}\|P_X b(\hat{\boldsymbol{\alpha}})\|^2 \le \frac{1}{\mu}\|b(\hat{\boldsymbol{\alpha}})\|^2 \le \frac{\lambda^2}{\nu^2\mu}\|L\boldsymbol{\alpha}\|^2. \qquad \text{(A.4)}
\end{aligned}$$

By Lemma 2 and Schur complement, the covariance matrix of $\hat{\boldsymbol{\beta}}$ is

$$\begin{aligned}
\mathrm{Var}(\hat{\boldsymbol{\beta}}) &\preceq \sigma^2(X^TX)^{-1} + \sigma^2(X^TX)^{-1}X^T(P_{X^\perp} + \lambda L)^{-1}X(X^TX)^{-1} \qquad \text{(A.5)} \\
&\preceq \sigma^2(X^TX)^{-1} + \frac{\sigma^2}{\nu}(X^TX)^{-1}X^TX(X^TX)^{-1} = \sigma^2(\frac{1}{\nu}+1)(X^TX)^{-1}.
\end{aligned}$$

Combining the squared bias (A.4) and variance (A.5) gives the bound (3.3) on MSE($\hat{\boldsymbol{\beta}}$). The mean squared prediction error can be similarly derived. With $\hat{\boldsymbol{V}} = \tilde{X}\hat{\boldsymbol{\theta}}$, we have

$$b(\hat{\boldsymbol{V}}) = \tilde{X} b(\hat{\boldsymbol{\theta}}) = -\lambda\tilde{X}(\tilde{X}^T\tilde{X} + \lambda M)^{-1}M\boldsymbol{\theta},$$

and

$$\mathrm{Var}(\hat{\boldsymbol{V}}) = \sigma^2\tilde{X}(\tilde{X}^T\tilde{X} + \lambda M)^{-1}\tilde{X}^T\tilde{X}(\tilde{X}^T\tilde{X} + \lambda M)^{-1}\tilde{X}^T.$$

Thus

$$\begin{aligned}
\mathbb{E}\|\hat{\boldsymbol{V}} - \mathbb{E}\boldsymbol{Y}\|^2 &= \|b(\hat{\boldsymbol{V}})\|^2 + \mathrm{tr}(\mathrm{Var}(\hat{\boldsymbol{V}})) \\
&\le \lambda^2(L\boldsymbol{\alpha})^T(P_{X^\perp} + \lambda L)^{-1}(L\boldsymbol{\alpha}) + \sigma^2\mathrm{tr}(S_\lambda^T S_\lambda) \\
&\le \frac{\lambda^2}{\nu}\|L\boldsymbol{\alpha}\|^2 + \sigma^2\|S_\lambda\|_F^2.
\end{aligned}$$

This completes the proof of Theorem 3.1. □

Proof of Proposition 2. Let $\tau_i$ be the $i$th largest eigenvalue of $L$ with the associated eigenvector $\boldsymbol{u}_k$. Assume $\boldsymbol{\alpha} = \sum_{i=n-k+1}^{n} \rho_i \boldsymbol{u}_i$. Without loss of generality, assume $\|\boldsymbol{\alpha}\|^2 = n$ thus $\sum_{i=n-k+1}^{n} \rho_i^2 = n$. In this situation, we need

$$\|L\boldsymbol{\alpha}\|^2 = \sum_{i=n-k+1}^{n} \rho_i^2 \tau_i^2 \le n^c.$$

Since $\sum_{i=n-k+1}^{n} \rho_i^2 \tau_i^2 \le \tau_{n-k+1}^2 \sum_{i=n-k+1}^{n} \rho_i^2 = n\tau_{n-k+1}^2$, it is sufficient to have $\rho_{n-k+1}^2 \le n^{-(1-c)}$. By basic graph spectral theory (Edwards, 2013), we can see that all of the eigenvalues of the lattice network can be written as

$$4\sin^2(\frac{\pi}{2}\frac{i}{\sqrt{n}}) + 4\sin^2(\frac{\pi}{2}\frac{j}{\sqrt{n}})$$

for some $(i, j) \in [\sqrt{n}] \times [\sqrt{n}]$. Thus it is sufficient to ensure

$$4\sin^2(\frac{\pi}{2}\frac{i}{\sqrt{n}}) + 4\sin^2(\frac{\pi}{2}\frac{j}{\sqrt{n}}) \le 4(\frac{\pi}{2}\frac{i}{\sqrt{n}})^2 + 4(\frac{\pi}{2}\frac{j}{\sqrt{n}})^2 \le n^{-\frac{1-c}{2}}.$$

For reasonably large $n$, the proportion of pairs $(i, j)$ satisfying the condition in $[\sqrt{n}] \times [\sqrt{n}]$ is approximately the area ratio between a $1/4$ sphere and a square, which is $\frac{1}{4\pi}\frac{n^{\frac{1+c}{2}}}{n}$. Therefore, the number of eigenvalues that satisfies the requirement is at least $Cn^{\frac{1+c}{2}}$ for some constant $C$.

□

For the easiness of comparison, we also give similar error bounds for the linear null model estimate, which is obtained as

$$\tilde{\boldsymbol{\theta}} = (\tilde{\boldsymbol{\alpha}}, \tilde{\boldsymbol{\beta}}) = \text{argmin}_{\boldsymbol{\alpha},\boldsymbol{\beta}} \|\boldsymbol{Y} - X\boldsymbol{\beta} - \boldsymbol{\alpha}\|^2 + \lambda\|\boldsymbol{\alpha}\|^2 \tag{A.6}$$

The following proposition shows that in the case of linear regression, the null model gives the same estimate of $\boldsymbol{\beta}$ as OLS.

Lemma 3. Let $\tilde{\boldsymbol{\beta}}$ be the estimate from null model. Then we have

$$\tilde{\boldsymbol{\beta}} = \hat{\boldsymbol{\beta}}_{OLS} = (X^T X)^{-1} X^T Y.$$

As a result, we have $\tilde{\boldsymbol{\alpha}} = \frac{1}{1+\lambda}(Y - X\hat{\boldsymbol{\beta}}_{OLS})$. Moreover, the estimation errors for the null model satisfy

$$\begin{aligned}\mathrm{MSE}(\tilde{\boldsymbol{\alpha}}) &= \|\boldsymbol{\alpha} - \frac{1}{1+\lambda}P_{X^\perp}\boldsymbol{\alpha}\|^2 + \frac{(n-p)\sigma^2}{(1+\lambda)^2} \\ &\le \frac{\lambda^2}{(1+\lambda)^2}\|\boldsymbol{\alpha}\|^2 + \frac{1}{(1+\lambda)^2}\|P_{X^\perp}\boldsymbol{\alpha}\|^2 + \frac{(n-p)\sigma^2}{(1+\lambda)^2}, \\ \mathrm{MSE}(\tilde{\boldsymbol{\beta}}) &= \|(X^TX)^{-1}X^T\boldsymbol{\alpha}\|^2 + \sigma^2\mathrm{tr}((X^TX)^{-1}), \\ \mathbb{E}\|\tilde{Y} - \mathbb{E}Y\|^2 &= \frac{\lambda^2}{(1+\lambda)^2}\|P_{X^\perp}\boldsymbol{\alpha}\|^2 + \sigma^2(p + \frac{n-p}{(1+\lambda)^2}).\end{aligned}$$

In particular, the optimal MSPE is

$$\mathbb{E}\|\tilde{Y} - \mathbb{E}Y\|^2 = \frac{(n-p)\sigma^2\|P_{X^\perp}\boldsymbol{\alpha}\|^2}{(n-p)\sigma^2 + \|P_{X^\perp}\boldsymbol{\alpha}\|^2}$$

which is achieved when $\lambda = \frac{(n-p)\sigma^2}{\|P_{X^\perp}\boldsymbol{\alpha}\|^2}$.

Proof of Lemma 3. Notice that $X$ is column centered, so we always have $\mathbf{1}^TX = 0$, which ensures

$$\hat{\boldsymbol{\beta}}_{OLS} = (X^TX)^{-1}X^TY.$$

The solution of the null model is given by

$$\begin{bmatrix}\tilde{\boldsymbol{\alpha}} \\ \tilde{\boldsymbol{\beta}}\end{bmatrix} = \begin{bmatrix}(1+\lambda)I_n & X \\ X^T & X^TX\end{bmatrix}^{-1}\begin{bmatrix}Y \\ X^TY\end{bmatrix}. \tag{A.7}$$

By block matrix inverse formula, we have

$$\begin{aligned}\tilde{\boldsymbol{\beta}} &= -\frac{1+\lambda}{\lambda}\frac{1}{1+\lambda}(X^TX)^{-1}X^TY + \frac{1+\lambda}{\lambda}(X^TX)^{-1}X^TY \\ &= (X^TX)^{-1}X^TY = \hat{\boldsymbol{\beta}}_{OLS}.\end{aligned}$$

The formula for $\tilde{\boldsymbol{\alpha}}$ and all the error bounds can then be obtained similarly as in Theorem 3.1. □

Proof of Theorem 3.3. Denote $\ell(\boldsymbol{\alpha} + X\boldsymbol{\beta}; \boldsymbol{Y})$ by $\ell(\boldsymbol{\theta})$. Define

$$M = \begin{bmatrix}L & 0_{n\times p} \\ 0_{p\times n} & 0_{p\times p}\end{bmatrix}.$$

The matrix $M^*$ is defined similarly. Then by the optimality of $\hat{\boldsymbol{\theta}}^*$ under $f^*$, we have

$$\begin{aligned}\ell(\hat{\boldsymbol{\theta}}^*) + \lambda\hat{\boldsymbol{\theta}}^{*T} M^*\hat{\boldsymbol{\theta}}^* &= f^*(\hat{\boldsymbol{\theta}}^*) \\ &\le f^*(\hat{\boldsymbol{\theta}}) \\ &= \ell(\hat{\boldsymbol{\theta}}) + \lambda\hat{\boldsymbol{\theta}}^T M^*\hat{\boldsymbol{\theta}} \\ &\le \ell(\hat{\boldsymbol{\theta}}) + \lambda(1+\epsilon)\hat{\boldsymbol{\theta}}^T M\hat{\boldsymbol{\theta}},\end{aligned} \tag{A.8}$$

in which the last inequality can be easily derived from (3.7) by noticing that $M^*$ has all zeros except in the upper left corner. By Taylor expansion of $\ell$ at $\hat{\boldsymbol{\theta}}$, we have

$$\begin{aligned}\ell(\hat{\boldsymbol{\theta}}^*) &= \ell(\hat{\boldsymbol{\theta}}) + \bigtriangledown\ell(\hat{\boldsymbol{\theta}})^T(\hat{\boldsymbol{\theta}}^* - \hat{\boldsymbol{\theta}}) + \frac{1}{2}(\hat{\boldsymbol{\theta}}^* - \hat{\boldsymbol{\theta}})^T \bigtriangledown^2 \ell(\bar{\boldsymbol{\theta}})(\hat{\boldsymbol{\theta}}^* - \hat{\boldsymbol{\theta}}) \\ &= \ell(\hat{\boldsymbol{\theta}}) + \bigtriangledown\ell(\hat{\boldsymbol{\theta}})^T(\hat{\boldsymbol{\theta}}^* - \hat{\boldsymbol{\theta}}) + \frac{1}{2}(\hat{\boldsymbol{\theta}}^* - \hat{\boldsymbol{\theta}})^T(\bigtriangledown^2\ell(\bar{\boldsymbol{\theta}}) + 2\lambda M)(\hat{\boldsymbol{\theta}}^* - \hat{\boldsymbol{\theta}}) \\ &\qquad - \lambda(\hat{\boldsymbol{\theta}}^* - \hat{\boldsymbol{\theta}})^T M(\hat{\boldsymbol{\theta}}^* - \hat{\boldsymbol{\theta}}) \\ &\ge \ell(\hat{\boldsymbol{\theta}}) + \bigtriangledown\ell(\hat{\boldsymbol{\theta}})^T(\hat{\boldsymbol{\theta}}^* - \hat{\boldsymbol{\theta}}) + \frac{m}{2}\|\hat{\boldsymbol{\theta}}^* - \hat{\boldsymbol{\theta}}\|^2 - \lambda(\hat{\boldsymbol{\theta}}^* - \hat{\boldsymbol{\theta}})^T M(\hat{\boldsymbol{\theta}}^* - \hat{\boldsymbol{\theta}}).\end{aligned} \tag{A.9}$$

In (A.9), $\bar{\boldsymbol{\theta}}$ is some point between $\hat{\boldsymbol{\theta}}$ and $\hat{\boldsymbol{\theta}}^*$ and the last inequality comes from the strong convexity assumption on $f$. Substituting (A.9) into (A.8) yields

$$\begin{aligned}\frac{m}{2}\|\hat{\boldsymbol{\theta}}^* - \hat{\boldsymbol{\theta}}\|^2 &\le - \bigtriangledown \ell(\hat{\boldsymbol{\theta}})^T(\hat{\boldsymbol{\theta}}^* - \hat{\boldsymbol{\theta}}) + \lambda(\hat{\boldsymbol{\theta}}^* - \hat{\boldsymbol{\theta}})^T M(\hat{\boldsymbol{\theta}}^* - \hat{\boldsymbol{\theta}}) \\ &\qquad + \lambda(1+\epsilon)\hat{\boldsymbol{\theta}}^T M\hat{\boldsymbol{\theta}} - \lambda\hat{\boldsymbol{\theta}}^{*T} M^*\hat{\boldsymbol{\theta}}^* \\ &= - \bigtriangledown \ell(\hat{\boldsymbol{\theta}})^T(\hat{\boldsymbol{\theta}}^* - \hat{\boldsymbol{\theta}}) + \lambda(2+\epsilon)\hat{\boldsymbol{\theta}}^T M\hat{\boldsymbol{\theta}} \\ &\qquad + \lambda\hat{\boldsymbol{\theta}}^{*T} M\hat{\boldsymbol{\theta}}^* - \lambda\hat{\boldsymbol{\theta}}^{*T} M^*\hat{\boldsymbol{\theta}}^* - 2\lambda\hat{\boldsymbol{\theta}}^T M\hat{\boldsymbol{\theta}}^*.\end{aligned} \tag{A.10}$$

Since $\hat{\boldsymbol{\theta}}$ is the minimizer of $f$, we have the stationary condition

$$\bigtriangledown \ell(\hat{\boldsymbol{\theta}}) + 2\lambda M\hat{\boldsymbol{\theta}} = \mathbf{0}. \tag{A.11}$$

Substituting (A.11) into (A.10) gives

$$\begin{aligned}\frac{m}{2}\|\hat{\boldsymbol{\theta}}^* - \hat{\boldsymbol{\theta}}\|^2 &\le 2\lambda\hat{\boldsymbol{\theta}}^T M(\hat{\boldsymbol{\theta}}^* - \hat{\boldsymbol{\theta}}) + \lambda(2+\epsilon)\hat{\boldsymbol{\theta}}^T M\hat{\boldsymbol{\theta}} + \lambda\hat{\boldsymbol{\theta}}^{*T} M\hat{\boldsymbol{\theta}}^* \\ &\qquad - \lambda\hat{\boldsymbol{\theta}}^{*T} M^*\hat{\boldsymbol{\theta}}^* - 2\lambda\hat{\boldsymbol{\theta}}^T M\hat{\boldsymbol{\theta}}^* \\ &= \epsilon\lambda\hat{\boldsymbol{\theta}}^T M\hat{\boldsymbol{\theta}} + \lambda\hat{\boldsymbol{\theta}}^{*T} M\hat{\boldsymbol{\theta}}^* - \lambda\hat{\boldsymbol{\theta}}^{*T} M^*\hat{\boldsymbol{\theta}}^* \\ &\le \epsilon\lambda\hat{\boldsymbol{\theta}}^T M\hat{\boldsymbol{\theta}} + \epsilon\lambda\hat{\boldsymbol{\theta}}^{*T} M\hat{\boldsymbol{\theta}}^* \\ &\le \epsilon\lambda\hat{\boldsymbol{\theta}}^T M\hat{\boldsymbol{\theta}} + \frac{\epsilon}{1-\epsilon}\lambda\hat{\boldsymbol{\theta}}^{*T} M\hat{\boldsymbol{\theta}}^* \\ &= \epsilon\lambda\hat{\boldsymbol{\alpha}}^T L\hat{\boldsymbol{\alpha}} + \frac{\epsilon}{1-\epsilon}\lambda\hat{\boldsymbol{\alpha}}^{*T} L\hat{\boldsymbol{\alpha}}^*, \end{aligned} \tag{A.12}$$

where we use (3.7) again. This gives the bound we need. However, it would be better to have a bound with a dominant term that only depends on $\hat{\boldsymbol{\alpha}}$ and $L$. Thus we rearrange the terms as

$$\begin{aligned}\frac{m}{2}\|\hat{\boldsymbol{\theta}}^* - \hat{\boldsymbol{\theta}}\|^2 &\le \epsilon\lambda\hat{\boldsymbol{\alpha}}^T L\hat{\boldsymbol{\alpha}} + \frac{\epsilon}{1-\epsilon}\lambda\hat{\boldsymbol{\alpha}}^{*T} L\hat{\boldsymbol{\alpha}}^* \\ &\le \epsilon\lambda\hat{\boldsymbol{\alpha}}^T L\hat{\boldsymbol{\alpha}} + (1+2\epsilon)\lambda\hat{\boldsymbol{\alpha}}^{*T} L\hat{\boldsymbol{\alpha}}^* \\ &= \epsilon\lambda[2\hat{\boldsymbol{\alpha}}^T L\hat{\boldsymbol{\alpha}} + (\hat{\boldsymbol{\alpha}}^{*T} L\hat{\boldsymbol{\alpha}}^* - \hat{\boldsymbol{\alpha}}^T L\hat{\boldsymbol{\alpha}}) + 2\epsilon\hat{\boldsymbol{\alpha}}^{*T} L\hat{\boldsymbol{\alpha}}^*] \\ &\le \epsilon\lambda[2\hat{\boldsymbol{\alpha}}^T L\hat{\boldsymbol{\alpha}} + |\hat{\boldsymbol{\alpha}}^{*T} L\hat{\boldsymbol{\alpha}}^* - \hat{\boldsymbol{\alpha}}^T L\hat{\boldsymbol{\alpha}}| + 2\epsilon\hat{\boldsymbol{\alpha}}^{*T} L\hat{\boldsymbol{\alpha}}^*], \end{aligned} \tag{A.13}$$

in which the second inequality comes from the fact that $\frac{1}{1-\epsilon} < 1 + 2\epsilon$ for $\epsilon < 1/2$. Note that we expect $|\hat{\boldsymbol{\alpha}}^{*T} L\hat{\boldsymbol{\alpha}}^* - \hat{\boldsymbol{\alpha}}^T L\hat{\boldsymbol{\alpha}}|$ to be negligible compared to the first term.

We now proceed to proving the second bound that only involves $\|\hat{\boldsymbol{\alpha}}\|$. By Taylor expansion, we have, for any $\boldsymbol{\theta}, \boldsymbol{\theta}_0 \in \mathbb{R}^n$,

$$\begin{aligned} f^*(\boldsymbol{\theta}) &= f^*(\boldsymbol{\theta}_0) + \triangledown f^*(\boldsymbol{\theta}_0)^T(\boldsymbol{\theta} - \boldsymbol{\theta}_0) + \frac{1}{2}(\boldsymbol{\theta} - \boldsymbol{\theta}_0)^T \triangledown^2 f^*(\tilde{\boldsymbol{\theta}})(\boldsymbol{\theta} - \boldsymbol{\theta}_0) \\ &\ge f^*(\boldsymbol{\theta}_0) + \triangledown f^*(\boldsymbol{\theta}_0)^T(\boldsymbol{\theta} - \boldsymbol{\theta}_0) + \frac{m}{2}\|\boldsymbol{\theta} - \boldsymbol{\theta}_0\|^2, \end{aligned}$$

where the inequality follows from strong convexity. In particular, taking $\boldsymbol{\theta} = \hat{\boldsymbol{\theta}}$ and $\boldsymbol{\theta}_0 = \hat{\boldsymbol{\theta}}^*$ and noticing that $\triangledown f^*(\hat{\boldsymbol{\theta}}^*) = \mathbf{0}$, we get

$$\|\hat{\boldsymbol{\theta}}^* - \hat{\boldsymbol{\theta}}\|^2 \le \frac{2}{m}(f^*(\hat{\boldsymbol{\theta}}) - f^*(\hat{\boldsymbol{\theta}}^*)).$$

Strong convexity also implies (equation (9.9) of (Boyd and Vandenberghe, 2004)) that

$$(f^*(\hat{\boldsymbol{\theta}}) - f^*(\hat{\boldsymbol{\theta}}^*)) \le \frac{1}{2m}\|\triangledown f^*(\hat{\boldsymbol{\theta}})\|^2.$$

Combining the two parts, we have

$$\|\hat{\boldsymbol{\theta}}^* - \hat{\boldsymbol{\theta}}\|^2 \leq \frac{1}{m^2}\|\bigtriangledown f^*(\hat{\boldsymbol{\theta}})\|^2 = \frac{1}{m^2}\|\bigtriangledown f^*(\hat{\boldsymbol{\theta}}) - \bigtriangledown f(\hat{\boldsymbol{\theta}})\|^2, \tag{A.14}$$

in which the last equality comes from the fact that $\bigtriangledown f(\hat{\boldsymbol{\theta}}) = \mathbf{0}$. From (3.8), the gradients of $f$ and $f^*$ are

$$\bigtriangledown f(\hat{\boldsymbol{\theta}}) = \bigtriangledown \ell + 2\lambda M \hat{\boldsymbol{\theta}}, \quad \bigtriangledown f^*(\hat{\boldsymbol{\theta}}) = \bigtriangledown \ell + 2\lambda M^* \hat{\boldsymbol{\theta}}.$$

Thus the difference between $\hat{\boldsymbol{\theta}}^*$ and $\hat{\boldsymbol{\theta}}$ can be bounded by

$$\|\hat{\boldsymbol{\theta}}^* - \hat{\boldsymbol{\theta}}\|^2 \leq \frac{1}{m^2}\|2\lambda(M - M^*)\hat{\boldsymbol{\theta}}\|^2. \tag{A.15}$$

Finally, from (3.7), we obtain

$$\begin{aligned}
\|2\lambda(M - M^*)\hat{\boldsymbol{\theta}}\|^2 &= \|2\lambda(L - L^*)\hat{\boldsymbol{\alpha}}\|^2 \\
&\leq 4\lambda^2\|L - L^*\|_2^2\|\hat{\boldsymbol{\alpha}}\|^2 \\
&\leq 4\lambda^2\epsilon^2\|L\|_2^2\|\hat{\boldsymbol{\alpha}}\|^2. \qquad \text{(A.16)}
\end{aligned}$$

Combining (A.15) and (A.16) yields the second bound and completes the proof. □

## APPENDIX B: COMPLEXITY OF SOLVING RNC ESTIMATOR BY BLOCK ELIMINATION

We calculate the complexity of solving RNC estimator here assuming the block elimination strategy described in Section 2.6 is used. The first major part is solving an $n \times n$ sparse symmetric diagonal dominant system to obtain $(I + \lambda L)^{-1}X$ and $(I + \lambda L)^{-1}\boldsymbol{b}_1$ in the estimator. Using the linear system notations, we want to solve

$$A\boldsymbol{x} = \boldsymbol{b}$$

where $A = I + \lambda L$. Naively solve it by Cholesky decomposition ignoring special structures would result in $O(n^3)$ operations. When $A$ is sparse as in a great many of applications, we can first find a permutation matrix $P$ to permute $A$ and then find sparse factorization for the resulting permuted matrix

$$PAP^T = LL^T.$$

The operation counts in this step depends on the heuristic algorithm to find a good permutation, the number of nonzero elements in $A$ (which is $n + 2|E|$ in our setting) and the positions of these nonzeros (depicted by the network). Roughly speaking, it depends on $\sum_i d_i^2$ (Spielman, 2010). Though the general complexity is not available, it is shown in Lipton, Rose and Tarjan (1979) that the complexity for the network transformed from a $\sqrt{n} \times \sqrt{n}$ grid is $O(n^{3/2})$ by using an algorithm called George's Nested Dissection. Solving both $(I + \lambda L)^{-1}X$ and $(I + \lambda L)^{-1}\boldsymbol{b}_1$ thus requires $O(n^{3/2} + pn)$ and when $n$ dominates $p$, we just have $O(n^{3/2})$ there. We refer readers to Lipton, Rose and Tarjan (1979) for details.

Alternatively, one can solve the system approximately by iterative methods (Spielman, 2010; Koutis, Miller and Peng, 2010). In particular, Koutis, Miller and Peng (2010) propose an iterative algorithm with preconditioning such that for any $n$-node network, an approximate solution $\hat{\boldsymbol{x}}$ of accuracy

$$\|\hat{\boldsymbol{x}} - A^{-1}\boldsymbol{b}\|_A < \epsilon\|A^{-1}\boldsymbol{b}\|_A$$

can be computed in expected time $O(m \log^2 n \log(1/\epsilon))$ where $m = n + 2|E|$ and the $A$-norm is defined by

$$\|\boldsymbol{x}\|_A = \sqrt{\boldsymbol{x}^T A \boldsymbol{x}}.$$

To solve both $(I + \lambda L)^{-1}X$ and $(I + \lambda L)^{-1}\boldsymbol{b}_1$, this is expected to takes $O(pm \log^2 n \log(1/\epsilon))$ operations. Notice that even if $A$ is fully dense with $n^2$ nonzero entries, the cost is still much lower than the naive solving.

The rest steps in the block elimination only involve matrix multiplications and general solving for a $p \times p$ symmetric positive definite system. The order is then $O(np^2 + p^3)$, the same as OLS procedure.

In summary, if one tries to compute the estimator exactly, the order depends on the network connecting the samples. When the network is from a $\sqrt{n} \times \sqrt{n}$ grid, the complexity is in the order of $O(n^{3/2} + pn + np^2 + p^3)$. If approximate methods are used instead, the order is expected to be $O(p(n + 2|E|) \log^3 n + np^2 + p^3)$ for general networks with high accuracy (taking approximation tolerance $\epsilon = O(1/n)$).

Both of dense and sparse Cholesky factorizations can be further parallelized on modern distributed systems (Bosilca et al., 2012; Faverge and Ramet, 2008; Lacoste et al., 2014), when high computational performance is needed. The complexity in such settings heavily depends the systems.

## APPENDIX C: COEFFICIENTS OF RECREATIONAL ACTIVITY LINEAR MODELS

In the example of Section 5.1, we use linear regression to predict recreational activity level from nine demographic covariates. The covariate coefficients from OLS and RNC regressions are shown in Table 4. Most of the coefficients are similar for the two models, suggesting that most of the variables do not contain (or mask) network structural information. The only covariate that is relatively significant in OLS but has a substantially smaller effect in RNC is the indicator variable "race-black". This suggests that race follows a network cohesion pattern, and thus is not as important for RNC since it is already getting network information elsewhere.

TABLE 4
*Estimated covariate coefficients from OLS and RNC linear regression on the recreational activity example. The p-values are for the OLS estimate.*

| category (contrast) | covariate | OLS | RNC | p-value (OLS) |
|---|---|---|---|---|
| | age | -0.086 | -0.088 | 0.065 |
| sex (male) | female | 0.229 | 0.241 | 0.003 |
| grade (other) | grade11to12 | 0.206 | 0.212 | 0.078 |
| race (other) | white | 0.023 | -0.029 | 0.733 |
| | black | 0.539 | 0.426 | 0.007 |
| | Asian | 0.346 | 0.512 | 0.081 |
| | native | 0.369 | 0.252 | 0.407 |
| born in U.S. (no) | yes | -0.059 | 0.090 | 0.290 |
| living with mother (no) | yes | 0.095 | 0.162 | 0.251 |
| living with father (no) | yes | -0.089 | -0.047 | 0.620 |
| parents in labor market (no) | yes | -0.193 | 0.018 | 0.957 |
| mother education (no high school) | high school | -0.039 | -0.021 | 0.861 |
| | more than high school | -0.116 | -0.000 | 1.000 |
| | college | 0.108 | 0.163 | 0.226 |
| | unknown | -0.061 | -0.051 | 0.681 |
| father education (no high school) | high school | -0.132 | -0.127 | 0.326 |
| | more than high school | 0.012 | -0.040 | 0.814 |
| | college | -0.049 | -0.026 | 0.853 |
| | unknown | -0.330 | -0.336 | 0.006 |

## APPENDIX D: SENSITIVITY TO MISSING DATA

The AddHealth data set contains many records with missing values, and we used imputation in both examples to handle the missing data. Here we report results of a sensitivity analysis to the amount of missing data.

In the recreational activity example, we remove an additional fraction $p_m$ of records in each column at random, where $p_m$ varies from 0 to 0.5; if the original column was missing $m$ values, it will now be missing $m(1 + p_m)$ records. When $p_m = 0$, the results match the ones reported in Section 4. Table 5 shows the corresponding RMSEs for the full model with all 10 predictors, calculated in the same way as in Section 4, for a range of values of $p_m$. The relative rankings of the five models never change, although there are some small numerical changes in the errors. This very robust performance suggests that our results are not sensitive to proportion of missing data.

TABLE 5

*Prediction errors of five models with missing data imputation, with varying proportion of additional missing values. All other columns are compared with RNC by a paired two-sample t-test. ** indicates a p-value* $\leq 10^{-4}$ *and * indicates a p-value* $\in (10^{-4}, 10^{-2})$.

| $p_m$ | OLS & Null | Boosting | SIM | RNC | RNC-LA | oracle-Bayes |
|---|---|---|---|---|---|---|
| 0% | 1.219 ** | 1.210 ** | 1.188 ** | 1.163 | 1.176 * | 1.175 * |
| 10% | 1.219 ** | 1.211 ** | 1.186 ** | 1.163 | 1.174 * | 1.171 |
| 20% | 1.216 ** | 1.210 ** | 1.184 ** | 1.160 | 1.172 * | 1.169 * |
| 30% | 1.218 ** | 1.213 ** | 1.188 ** | 1.164 | 1.174 * | 1.174 * |
| 40% | 1.220 ** | 1.213 ** | 1.198 ** | 1.167 | 1.186 ** | 1.179 * |
| 50% | 1.216 ** | 1.210 ** | 1.185 ** | 1.163 | 1.175 * | 1.172 |

In the marijuana usage example, we conduct the same experiment. However, the number of records with missing values is much smaller in the home-survey data (used for marijuana example) than the school survey data (used in the recreational activity example). Therefore, we take a larger range for $p_m$ from 0 to 2. The results of the corresponding prediction iAUC for the four models (with all the five variables) are shown in Table 6. Again, the relative ranking is the same as in Section 4 for all different values of $p_m$. Moreover, the iAUCs are also very stable across different settings of $p_m$.

TABLE 6

*Average integrated AUC (iAUC) for survival prediction ROC curves for age 14-17 with artificially increased missing values (by $p_m$). The average is taken over 50 random splits of the data into 60 test samples and 587 training samples. All values are compared with the columns of RNC by a paired two-sample t-test. ** indicates a p-value $\leq 10^{-4}$ and * indicates a p-value $\in (10^{-4}, 10^{-2})$.*

| $p_m$ | Cox & Null | Boosting | SIM | RNC | RNC-LA |
|---|---|---|---|---|---|
| 0 | 0.727 ** | 0.742 * | 0.743 | 0.748 | 0.766 ** |
| 0.5 | 0.727 ** | 0.742 * | 0.743 | 0.748 | 0.766 ** |
| 1 | 0.726 ** | 0.740 * | 0.742 | 0.747 | 0.765 ** |
| 1.5 | 0.726 ** | 0.740 * | 0.742 | 0.747 | 0.765 ** |
| 2 | 0.729 ** | 0.744 * | 0.745 | 0.749 | 0.767 ** |

# REFERENCES

Abbe, E. (2017). Community detection and stochastic block models: recent developments. *arXiv preprint arXiv:1703.10146*.

Amini, A. A., Chen, A., Bickel, P. J. and Levina, E. (2013). Pseudo-likelihood methods for community detection in large sparse networks. *The Annals of Statistics* **41** 2097–2122.

Asur, S. and Huberman, B. A. (2010). Predicting the future with social media. In *Web Intelligence and Intelligent Agent Technology (WI-IAT), 2010 IEEE/WIC/ACM International Conference on* **1** 492–499. IEEE.

Belkin, M. and Niyogi, P. (2003). Laplacian eigenmaps for dimensionality reduction and data representation. *Neural Computation* **15** 1373–1396.

Belkin, M., Niyogi, P. and Sindhwani, V. (2006). Manifold regularization: A geometric framework for learning from labeled and unlabeled examples. *Journal of Machine Learning Research* **7** 2399–2434.

Bengio, Y., Paiement, J.-F., Vincent, P., Delalleau, O., Le Roux, N. and Ouimet, M. (2004). Out-of-sample extensions for lle, isomap, mds, eigenmaps, and spectral clustering. *Advances in Neural Information Processing Systems* **16** 177–184.

Besag, J. (1974). Spatial interaction and the statistical analysis of lattice systems. *Journal of the Royal Statistical Society. Series B (Methodological)* 192–236.

Binkiewicz, N., Vogelstein, J. T. and Rohe, K. (2017). Covariate-assisted spectral clustering. *Biometrika* **104** 361-377.

Bosilca, G., Bouteiller, A., Danalis, A., Herault, T., Lemarinier, P. and Dongarra, J. (2012). DAGuE: A generic distributed DAG engine for high performance computing. *Parallel Computing* **38** 37–51.

Boyd, S. and Vandenberghe, L. (2004). *Convex optimization*. Cambridge University Press.

Bramoullé, Y., Djebbari, H. and Fortin, B. (2009). Identification of peer effects through social networks. *Journal of Econometrics* **150** 41–55.

Bühlmann, P. and Hothorn, T. (2007). Boosting algorithms: Regularization, prediction and model fitting. *Statistical Science* 477–505.

Cai, D., He, X. and Han, J. (2007). Spectral regression: A unified approach for sparse subspace learning. In *Seventh IEEE International Conference on Data Mining (ICDM 2007)* 73–82. IEEE.

Chaudhuri, K., Graham, F. C. and Tsiatas, A. (2012). Spectral Clustering of Graphs with General Degrees in the Extended Planted Partition Model. In *COLT* **23** 35–1.

Choi, D. (2017). Estimation of monotone treatment effects in network experiments. *Journal of the American Statistical Association* 1–9.

Christakis, N. A. and Fowler, J. H. (2007). The spread of obesity in a large social network over 32 years. *New England Journal of Medicine* **357** 370–379.

Cohen, M. B., Kyng, R., Miller, G. L., Pachocki, J. W., Peng, R., Rao, A. B. and Xu, S. C. (2014). Solving SDD linear systems in nearly m log 1/2 n time. In *Proceedings of the 46th Annual ACM Symposium on Theory of Computing* 343–352. ACM.

Cox, D. R. (1972). Regression models and life-tables. *Journal of the Royal Statistical Society. Series B (Methodological)* 187–220.

Cressie, N. (1990). The origins of kriging. *Mathematical geology* **22** 239–252.

Edwards, T. (2013). The discrete Laplacian of a rectangular grid.

Faverge, M. and Ramet, P. (2008). Dynamic scheduling for sparse direct solver on NUMA architectures. In *PARA'08*.

Fujimoto, K. and Valente, T. W. (2012). Social network influences on adolescent

substance use: disentangling structural equivalence from cohesion. *Social Science & Medicine* **74** 1952–1960.

Goldenberg, A., Zheng, A. X., Fienberg, S. E. and Airoldi, E. M. (2010). A survey of statistical network models. *Foundations and Trends® in Machine Learning* **2** 129–233.

Hallac, D., Leskovec, J. and Boyd, S. (2015). Network lasso: Clustering and optimization in large graphs. In *Proceedings of the 21th ACM SIGKDD international conference on knowledge discovery and data mining* 387–396. ACM.

Harris, K. M. (2009). *The National Longitudinal Study of Adolescent to Adult Health (Add Health), Waves I & II, 1994-1996; Wave III, 2001-2002; Wave IV, 2007–009 [machine-readable data file and documentation]*. Carolina Population Center, University of North Carolina at Chapel Hill.

Haynie, D. L. (2001). Delinquent peers revisited: Does network structure matter? *American Journal of Sociology* **106** 1013–1057.

Henderson, C. R. (1953). Estimation of variance and covariance components. *Biometrics* **9** 226–252.

Hothorn, T., Buehlmann, P., Kneib, T., Schmid, M. and Hofner, B. (2018). mboost: Model-Based Boosting R package version 2.9-0.

Kim, S., Pan, W. and Shen, X. (2013). Network-based penalized regression with application to genomic data. *Biometrics* **69** 582–593.

Kolaczyk, E. D. (2009). *Statistical Analysis of Network Data: Methods and Models*, 1st ed. Springer Publishing Company, Incorporated.

Koutis, I., Miller, G. L. and Peng, R. (2010). Approaching optimality for solving SDD linear systems. In *Foundations of Computer Science (FOCS), 2010 51st Annual IEEE Symposium on* 235–244. IEEE.

Lacoste, X., Faverge, M., Ramet, P., Thibault, S. and Bosilca, G. (2014). Taking advantage of hybrid systems for sparse direct solvers via task-based runtimes. In *Parallel & Distributed Processing Symposium Workshops (IPDPSW), 2014 IEEE International* 29–38. IEEE.

Land, S. R. and Friedman, J. H. (1997). Variable fusion: A new adaptive signal regression method Technical Report, Technical Report 656, Department of Statistics, Carnegie Mellon University Pittsburgh.

Le, C. M., Levina, E. and Vershynin, R. (2017). Concentration and regularization of random graphs. *Random Structures & Algorithms*.

Lee, L.-f. (2007). Identification and estimation of econometric models with group interactions, contextual factors and fixed effects. *Journal of Econometrics* **140** 333–374.

Lee, D. (2013). CARBayes: An R Package for Bayesian Spatial Modeling with Conditional Autoregressive Priors. *Journal of Statistical Software* **55** 1–24.

Li, T., Levina, E. and Zhu, J. (2016). netcoh: Statistical Modeling with Network Cohesion R package version 0.11.

Li, C. and Li, H. (2008). Network-constrained regularization and variable selection for analysis of genomic data. *Bioinformatics* **24** 1175–1182.

Li, C. and Li, H. (2010). Variable selection and regression analysis for graph-structured covariates with an application to genomics. *The Annals of Applied Statistics* **4** 1498.

Lin, X. (2010). Identifying peer effects in student academic achievement by spatial autoregressive models with group unobservables. *Journal of Labor Economics* **28** 825–860.

Lipton, R. J., Rose, D. J. and Tarjan, R. E. (1979). Generalized nested dissection. *SIAM Journal on Numerical Analysis* **16** 346–358.

Manski, C. F. (1993). Identification of endogenous social effects: The reflection problem. *The Review of Economic Studies* **60** 531–542.

MANSKI, C. F. (2013). Identification of treatment response with social interactions. *The Econometrics Journal* **16** S1–S23.

MICHELL, L. and WEST, P. (1996). Peer pressure to smoke: the meaning depends on the method. *Health Education Research* **11** 39–49.

NEWMAN, M. E. and CLAUSET, A. (2016). Structure and inference in annotated networks. *Nature Communications* **7**.

PAN, W., XIE, B. and SHEN, X. (2010). Incorporating predictor network in penalized regression with application to microarray data. *Biometrics* **66** 474–484.

PEARSON, M. and MICHELL, L. (2000). Smoke rings: social network analysis of friendship groups, smoking and drug-taking. *Drugs: Education, Prevention, and Policy* **7** 21–37.

PEARSON, M. and WEST, P. (2003). Drifting smoke rings. *Connections* **25** 59–76.

PHAN, T. Q. and AIROLDI, E. M. (2015). A natural experiment of social network formation and dynamics. *Proceedings of the National Academy of Sciences* **112** 6595–6600.

RADUCANU, B. and DORNAIKA, F. (2012). A supervised non-linear dimensionality reduction approach for manifold learning. *Pattern Recognition* **45** 2432–2444.

RAND, D. G., ARBESMAN, S. and CHRISTAKIS, N. A. (2011). Dynamic social networks promote cooperation in experiments with humans. *Proceedings of the National Academy of Sciences* **108** 19193–19198.

RUE, H. and HELD, L. (2005). *Gaussian Markov random fields: theory and applications.* CRC Press.

SADHANALA, V., WANG, Y.-X. and TIBSHIRANI, R. J. (2016). Graph Sparsification Approaches for Laplacian Smoothing. In *Proceedings of the 19th International Conference on Artificial Intelligence and Statistics* 1250–1259.

SEARLE, S. R., CASELLA, G. and MCCULLOCH, C. E. (2009). *Variance Components* **391**. John Wiley & Sons.

SHALIZI, C. R. and THOMAS, A. C. (2011). Homophily and contagion are generically confounded in observational social network studies. *Sociological Methods and Research* **40** 211–239.

SHARPNACK, J., SINGH, A. and KRISHNAMURTHY, A. (2013). Detecting activations over graphs using spanning tree wavelet bases. In *Artificial Intelligence and Statistics* 536–544.

SHI, J. and MALIK, J. (2000). Normalized cuts and image segmentation. *Pattern Analysis and Machine Intelligence, IEEE Transactions on* **22** 888–905.

SONG, X. and ZHOU, X.-H. (2008). A semiparametric approach for the covariate specific ROC curve with survival outcome. *Statistica Sinica* 947–965.

SPIELMAN, D. A. (2010). Algorithms, graph theory, and linear equations in Laplacian matrices. In *Proceedings of the International Congress of Mathematicians* **4** 2698–2722.

SPIELMAN, D. A. and TENG, S.-H. (2011). Spectral sparsification of graphs. *SIAM Journal on Computing* **40** 981–1025.

TENENBAUM, J. B., DE SILVA, V. and LANGFORD, J. C. (2000). A global geometric framework for nonlinear dimensionality reduction. *Science* **290** 2319–2323.

TIBSHIRANI, R., SAUNDERS, M., ROSSET, S., ZHU, J. and KNIGHT, K. (2005). Sparsity and smoothness via the fused lasso. *Journal of the Royal Statistical Society: Series B (Statistical Methodology)* **67** 91–108.

VAPNIK, V. (2013). *The Nature of Statistical Learning Theory.* Springer Science & Business Media.

VOGELSTEIN, J. T., RONCAL, W. G., VOGELSTEIN, R. J. and PRIEBE, C. E. (2013). Graph classification using signal-subgraphs: Applications in statistical connectomics. *Pattern Analysis and Machine Intelligence, IEEE Transactions on* **35** 1539–1551.

VURAL, E. and GUILLEMOT, C. (2016). Out-of-sample generalizations for supervised man-ifold learning for classification. *IEEE Transactions on Image Processing* **25** 1410–1424.

Wahba, G. et al. (1999). Support vector machines, reproducing kernel Hilbert spaces and the randomized GACV. *Advances in Kernel Methods-Support Vector Learning* **6** 69–87.

Waller, L. A. and Gotway, C. A. (2004). *Applied spatial statistics for public health data* **368**. John Wiley & Sons.

Wang, Y.-X., Sharpnack, J., Smola, A. and Tibshirani, R. J. (2016). Trend filtering on graphs. *Journal of Machine Learning Research* **17** 1–41.

Wolf, T., Schroter, A., Damian, D. and Nguyen, T. (2009). Predicting build failures using social network analysis on developer communication. In *Proceedings of the 31st International Conference on Software Engineering* 1–11. IEEE Computer Society.

Xu, Y., Dyer, J. S. and Owen, A. B. (2010). Empirical stationary correlations for semi-supervised learning on graphs. *The Annals of Applied Statistics* 589–614.

Yang, W., Sun, C. and Zhang, L. (2011). A multi-manifold discriminant analysis method for image feature extraction. *Pattern Recognition* **44** 1649–1657.

Zhang, Y., Levina, E., Zhu, J. et al. (2016). Community detection in networks with node features. *Electronic Journal of Statistics* **10** 3153–3178.

Zhou, D., Huang, J. and Schölkopf, B. (2005). Learning from labeled and unlabeled data on a directed graph. In *Proceedings of the 22nd international conference on Machine learning* 1036–1043. ACM.

Zhou, D., Bousquet, O., Lal, T. N., Weston, J. and Schölkopf, B. (2004). Learning with local and global consistency. In *Advances in Neural Information Processing Systems* 321–328.

1085 S. University Ave., Ann Arbor, MI 48109